\documentclass[journal]{IEEEtran}

\usepackage{cite}
\usepackage{url}
\usepackage{comment}

\ifCLASSINFOpdf
\else
\fi

\usepackage{graphicx} 

\usepackage{comment}
\usepackage[normalem]{ulem}

\usepackage{amsmath}
\usepackage[dvipsnames]{xcolor}

\usepackage{algorithm}
\usepackage{algorithmicx}
\usepackage{algpseudocode}

\usepackage{epstopdf}

\DeclareMathOperator*{\argmin}{arg\!\min}
\usepackage{mathtools}
\DeclarePairedDelimiter\floor{\lfloor}{\rfloor}
\ifCLASSOPTIONcompsoc
\usepackage[caption=false,font=normalsize,labelfont=sf,textfont=sf]{subfig}
\else
\usepackage[caption=false,font=footnotesize]{subfig}
\fi

\begin{document}
\title{Synchronous LoRa Communication by Exploiting Large-Area Out-of-Band Synchronization}

\author{\vspace{-25pt}}
\author{Luca~Beltramelli,~\IEEEmembership{Member,~IEEE,}
        Aamir~Mahmood,~\IEEEmembership{Senior~Member,~IEEE,}
        Paolo~Ferrari,~\IEEEmembership{Member,~IEEE,}
        Patrik~{\"O}sterberg,~\IEEEmembership{Member,~IEEE,}
        Mikael~Gidlund,~\IEEEmembership{Senior~Member,~IEEE,}
        and~Emiliano~Sisinni,~\IEEEmembership{Member,~IEEE}
\thanks{L.~Beltramelli, A.~Mahmood, P.~{\"O}sterberg, and M.~Gidlund  are with the Department of Information Systems and Technology, Mid Sweden University, 851~70~Sundsvall, Sweden, e-mail: luca.beltramelli@miun.se.

P.~Ferrari and E.~Sisinni are with the Department of Information Engineering, University
of Brescia, 25123 Brescia, Italy.}
\vspace{-25pt}}

\maketitle

\begin{abstract}
Many new narrowband low-power wide-area networks (LPWANs) (e.g., LoRaWAN, Sigfox) have opted to use pure ALOHA-like access for its reduced control overhead and asynchronous transmissions. Although asynchronous access reduces the energy consumption of IoT devices, the network performance suffers from high intra-network interference in dense deployments. Contrarily, adopting synchronous access can improve throughput and fairness, however, it requires time synchronization. Unfortunately, maintaining synchronization over the narrowband LPWANs wastes channel time and transmission opportunities. In this paper, we propose the use of out-of-band time-dissemination to relatively synchronize the LoRa devices and thereby facilitate resource-efficient slotted uplink communication. In this respect, we conceptualize and analyze a co-designed synchronization and random access communication mechanism that can effectively exploit technologies providing limited time accuracy, such as FM radio data system (\mbox{FM-RDS}). While considering the LoRa-specific parameters, we derive the throughput of the proposed mechanism,  compare it to a generic synchronous random access using in-band synchronization, and design the communication parameters under time uncertainty. We scrutinize the transmission time uncertainty of a device by introducing a clock error model that accounts for the errors in the synchronization source, local clock, propagation delay, and transceiver’s transmission time uncertainty. We characterize the time uncertainty of FM-RDS with hardware measurements and perform simulations to evaluate the proposed solution. The results, presented in terms of success probability, throughput, and fairness for a single-cell scenario, suggest that \mbox{FM-RDS}, despite its poor absolute synchronization, can be used effectively to realize time-slotted communication in LoRa with performance similar to that of more accurate time-dissemination technologies.
\end{abstract}
\vspace{-5pt}
\vspace{-5pt}
\IEEEpeerreviewmaketitle

\section{Introduction}
With the explosive growth of IoT applications, a new generation of wireless technologies has emerged: low-power wide-area networks (LPWANs)~\cite{vangelista2015long}. The design of the physical (PHY) and data link (DL) layers in LPWANs is driven by the requirements of IoT monitoring applications, where, instead of data rate, the focus in on energy-efficient, sporadic and mainly uplink communication over long distances. Simple distributed access methods are preferred since they scale well with the increasing number of devices and avoid the need for  scheduling or other control information. LPWANs operate in the sub-1GHz ISM band, which allows for better coverage and building penetration. The medium access control (MAC) is the most critical part of the DL layer; it seats just above the PHY layer, controlling how shared radio resources are utilized. For this reason, the MAC is one of the primary control knobs of any communication solution, in order to fully exploit the capabilities of the underlying PHY layer~\cite{MAC_design}.
In general, the MAC layer in LPWANs is based on a simple random access paradigm in which the devices can transmit their messages at any time, potentially leading to an increased interference.

Currently, the most diffused and widely accepted LPWAN technology is LoRaWAN~\cite{CR_LPWAN}, designed to support event-triggered asynchronous uplink communication in monitoring applications. LoRaWAN is based on a proprietary radio technology (LoRa), which employs an efficient chirp spread spectrum modulation. LoRa overcomes the poor bandwidth availability by using quasi-orthogonal virtual channels that offer different data rates according to the spreading factor (SF), a tunable parameter that also represents the number of bits coded per symbol~\cite{vangelista2015long}. In LoRaWAN, the devices transmit messages asynchronously using  pure ALOHA access mechanism. 
The limitation of pure ALOHA is revealed by the high collision probability affecting densely deployed LoRaWAN networks, thus limiting LoRaWAN scalability in large-scale deployments of IoT applications~\cite{bor2016lora}. 

In the literature, numerous works have investigated the suitability of alternative access mechanisms for improving the performance of LoRaWAN~\cite{9018210,polonelli2019slotted,to2018simulation}. In this sense, one possibility is to use a \textit{listen-before-talk} (LBT) mechanism. LBT is effectively used to limit the number of collisions in short-range wireless networks, whereas, the additional energy consumption due to overhearing and the wide coverage areas would lessen its gains in many LPWAN use-cases.
Alternatively, collisions can be reduced by using synchronous transmissions (random or scheduled), but this requires synchronization among the devices. Although synchronization could be provided directly using the in-band resources, it would sacrifice a portion of the already limited bandwidth available to LoRa, thus reducing the communication opportunities of the devices. In~\cite{haxhibeqiri2018low}, it is shown that the in-band synchronization overhead can reduce the message delivery rate of the devices, especially those using smaller SF. Alternatively, out-of-band synchronization, sometimes referred to as hardware-assisted synchronization~\cite{HWassisted}, could be preferable, even though a complementary (wireless) technology is required for receiving time-related information.

In this works, we propose a joint synchronization and communication scheme for synchronous uplink random access in LoRa networks, in which out-of-band synchronization events are tracked by the devices to relatively align their local clock times. This alignment allows to divide the time between two periodic synchronization events into timeslots for the synchronous random channel access. As an out-of-band synchronization source for our scheme, we investigate the use of the FM radio data system (FM-RDS) for its wide availability and indoor/outdoor coverage despite its limited synchronization accuracy. Without loss of generality on the synchronization source, our main contributions are: 
 \begin{enumerate}
    \item  Considering the characteristics of LoRa modulation, we design the parameters of the proposed time-slotted communication scheme (i.e., timeslot duration, guard times, and synchronization interval) for uniform and Gaussian distributed synchronization errors.  
    \item Based on an analytical model, we compare the throughput of synchronous random access in LoRa using out-of-band synchronization with that of a two-way in-band  message exchange synchronization (\cite{polonelli2019slotted, haxhibeqiri2018low}).
    \item Using the `propagation of uncertainty' principle, we study the cumulative effect of the timing errors affecting the transmissions of a device, including the uncertainties in the LoRa transceiver transmission time, propagation delay, and local clock error. The time uncertainty of FM-RDS is experimentally characterized and used to derive the local clock errors of a device.
    \item A discrete event simulator is used to determine the throughput, fairness, and transmission success probability of the proposed  synchronization and communication scheme in a single-cell scenario. The simulator incorporates the power-capture effects of LoRa modulation and the timing errors of the proposed synchronization scheme.
\end{enumerate}
The rest of the paper is organized as follows. Sec.~\ref{sec_related_works} presents related works, and discusses suitable out-of-band synchronization technologies. The proposed communication scheme is described in Sec.~\ref{TheProposedApproach}. Sec.~\ref{sec:th_comparison} compares the theoretical throughput of LoRa for out-of-band and in-band synchronization. The design of the synchronization and communication parameters is presented in Sec.~\ref{sec:model}, while its timing errors are analyzed in Sec.~\ref{TimingErrors}. Simulations results are presented in Sec.~\ref{Results}, and the conclusions are drawn in Sec.~\ref{Conclusion}.
\section{Background and Related Works}
\label{sec_related_works}
The performance of the uplink communication in a single channel LoRaWAN networks has been studied extensively in the literature using analytical models and simulations~\cite{9018210, rahmadhani2018lorawan, haxhibeqiri2017lora}. The results of these analyses have shown an exponential decrease in the coverage probability for an increasing number of devices in the network. Researchers have proposed power control~\cite{8449737} and SFs allocation algorithms~\cite{8115779} to increase the scalability and reliability of LoRa. However, the pure ALOHA channel access remains a major bottleneck to the performance of LoRaWAN in dense scenarios. Consequently, the use of more advanced random and scheduled access has been proposed for LoRa networks.

In~\cite{pham2018robust, liando2019known, to2018simulation}, the use of an LBT mechanism in LoRa networks is investigated; however, LBT is not particularly suited for the wide-area communication in LPWANs. Other works have studied the use of slotted ALOHA~\cite{polonelli2019slotted, 8848387, zorbas2020design}, which in theory, could reduce the interference in the network and consequently double the channel capacity of LoRa. However, providing the  synchronization required by slotted ALOHA to a high number of devices can be challenging due to the constraints dictated by limited resource availability (energy, bandwidth, computation capability) and unreliable links. This challenge is also shared by the scheduled channel access mechanisms that have been proposed for LoRa~\cite{7803607, piyare2018demand,haxhibeqiri2018low, 8591568, zorbas2020ts}.

Additionally, since most LPWAN applications do not require strict time synchronization, no standardized solutions for time-dissemination are provided. For instance, in LoRaWAN, only Class-B devices use synchronized downlink feature based on periodic beacons trasnmitted by the gateway. 
For this reason, different in-band synchronization methods have been investigated for LoRaWAN, including periodic beacons~\cite{reynders2018improving} two-way message exchange~\cite{polonelli2019slotted, haxhibeqiri2018low} and posteriori~\cite{SyncUncertainty} synchronization schemes. 
These studies have shown that accuracy in the order of few tens of milliseconds can be achieved \cite{polonelli2019slotted}, however, in-band synchronization reduces the transmission opportunities of the devices~\cite{haxhibeqiri2018low}, as detailed further in Sec.~\ref{sec:th_comparison}.
Given the limitations of the in-band synchronization, it is important to investigate the use of out-of-band time-dissemination technologies for synchronous communication in LoRaWAN.

\subsection{Out-of-band Time-Dissemination Technologies}
Out-of-band synchronization requires to receive synchronization information on a separate channel from data traffic. A practical approach to realize out-of-band synchronization is to use one of the available time-dissemination technologies, however, not all time-dissemination technologies are equally suited for use in LPWANs.
To establish time synchronization in wide-area wireless networks, receivers offering very good sensitivity are required to achieve wide-area coverage complemented by indoor/outdoor capability. Moreover, the whole system must be designed to maximize the accuracy of the time of arrival estimation~\cite{ToA_Timestamp}. Examples of large-area time-dissemination technologies that can be exploited include Global Navigation Satellite Systems (e.g., GPS~\cite{GPSAccuracy}), radio-controlled clocks (e.g., DCF77~\cite{DCF_SDR}), and FM radio data broadcasting (e.g.,  \mbox{FM-RDS}~\cite{RDS}). 

The use of GPS and DCF77 has been previously proposed in \cite{8703036} to enable LoRa Mesh communication. In this article, the \mbox{FM-RDS} is considered as: a) it offers better reception both indoor and outdoor, b) it ensures reliability due to the availability of many FM stations in a single region, and c) it is compatible with low-power applications~\cite{RDSConsumption,6987310}. Inherently, FM-RDS offers  absolute synchronization service, based on the periodic (every minute) transmissions of the clock time and date (CT)-group messages, containing a timestamp. Unfortunately, the accuracy is poor, in the order of 100 ms~\cite{RDS}, and broadcasting stations are not required to be UTC synchronized. On the contrary, if relative  synchronization is considered, accuracy in the order of 100s of microseconds~\cite{RDSSynch} is feasible. 

\begin{figure*}[!htp]
	\centering
	\includegraphics[width=0.65\linewidth]{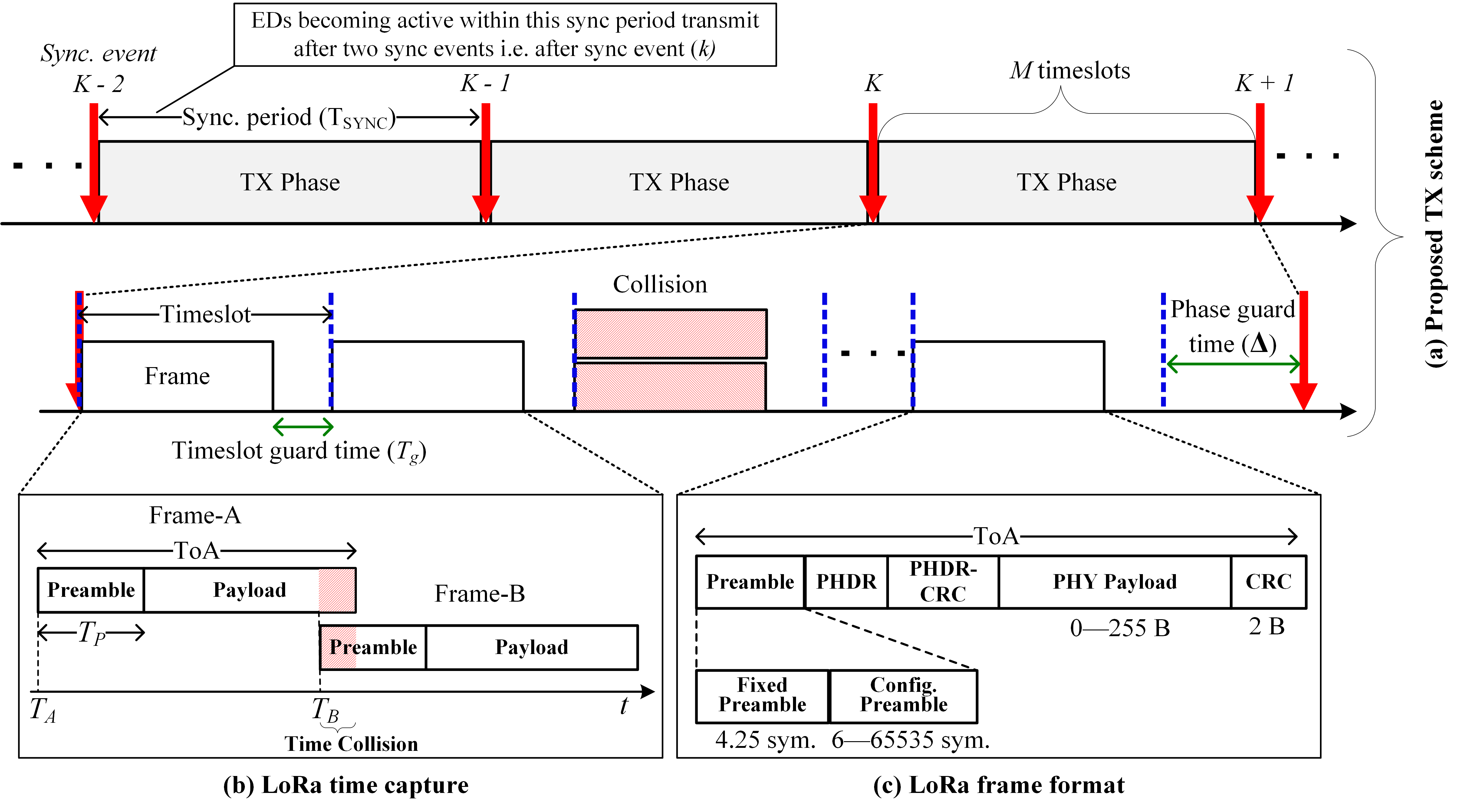}
	\caption{Proposed scheme: (a) Structure of the timeslots in the proposed TX phase; (b) Collision of LoRa frames and (c) Structure of a LoRa data frame.}
	\label{fig:frame}
	\vspace{-10pt}
\end{figure*}

\section{Exploiting Out-of-Band Time-Dissemination for Synchronous LoRa Communication}
\label{TheProposedApproach}
This section presents the proposed synchronous communication for LoRa, exploiting periodic out-of-band events for the relative synchronization of the devices.
The scheme is generic and can be used with any out-of-band time-dissemination technology discussed in the previous section. 
Hereafter, we outline the assumptions used in the article before describing the design of the scheme. 

\subsection{Assumptions}
{We consider a LoRa network consisting of a fixed number of LoRa end devices (EDs) and a single  gateway (GW).} All EDs operate on the same channel, under the regulatory duty-cycle constraints, limiting their activity factor $\alpha$. Following the previous analyses of the LoRa modulation in the literature~\cite{101007, 7803607}, we assume that SFs are perfectly orthogonal, allowing to consider each SF as an independent virtual channel with power-capture. Leveraging on the orthogonality, we only consider EDs using the same SF in the model. Moreover, we assume that all transmitted messages have a fixed payload size, resulting in the same time-on-air (ToA). 
If EDs are allowed to transmit messages of different sizes, the slotted communication can still function, although at a lower channel utilization efficiency.  The duration of a timeslot is designed to allow for the transmission of a message plus a guard time ($T_g$), preventing collisions between transmissions in adjacent slots. 
We assume that all EDs are equipped with a low-power out-of-band receiver~\cite{RDSConsumption}, capable of detecting the periodically transmitted synchronization events, nominally every $T_\text{SYNC}$. Finally, since most of the intended monitoring applications rely on uplink-only communication, collisions with downlink traffic are neglected.

\subsection{The Proposed Communication Scheme}
\label{sec:Sync_alg}
In the proposed scheme, the EDs access the channel using a time-slotted protocol, which requires all active EDs to be relatively synchronized.
Devices are assumed to transmit sporadically, hence, a random access to the channel is adopted instead of a scheduled access. Nonetheless, more advanced scheduling techniques, such as the one proposed in \cite{haxhibeqiri2018low}, can be combined with the proposed communication scheme to further enhance the scalability of the proposed solution.
Because the synchronization is only required for transmitting a message and transmissions are limited by duty-cycle restrictions, we consider on-demand instead of continuous synchronization. On-demand synchronization requires the devices to be synchronized only when it is necessary for accessing the channel, reducing the energy consumption by limiting the amount of time the out-of-band receiver remains active. The  synchronization and communication scheme  is designed such that it is independent of the accuracy of the time information transmitted by the time-dissemination technology.
Algorithm~\ref{alg:ALG1} shows the pseudo-code with the steps required for an ED to synchronize and transmit a message to the GW. At this stage, we assume that the detection of the synchronization events is immune to timing errors, a detailed analysis of the timing errors affecting the synchronization in the proposed scheme is deferred to Sec.~\ref{TimingErrors}.

The proposed scheme consists of three phases: a sleep phase, a synchronization phase and a transmission phase (c.f., Fig.~\ref{fig:frame}.a). 
\subsubsection{Sleep Phase}
The devices are in sleep mode for most of their lifetime, becoming active only when they have a message for the gateway.
A device saves energy by remaining in sleep mode with both the out-of-band receiver and the in-band transceiver inactive until a message is generated. When a message is queued for transmission, the device becomes active and begins the synchronization phase by activating the out-of-band receiver.
\subsubsection{Synchronization Phase}
The purpose of the synchronization phase is to ensure that all active devices are relatively synchronized using the periodic broadcast messages of the out-of-band time-dissemination technology.
An active device wakes up its out-of-band receiver to receive the next two synchronization events.  When a synchronization event is detected, an interrupt is generated by the out-of-band receiver and the device local clock $C(t)$ is saved. 
Upon receiving the second synchronization event, the synchronization phase is complete, and the device estimates its local clock rate $\beta$ as the ratio between the number of clock periods $T_0$ counted between the arrival time of the last two synchronization events at $t_{-1}$ and $t_0$, respectively, and the nominal synchronization period $T_\text{SYNC}$. Each active device keeps track of the time during the subsequent transmission phase from the arrival of the last synchronization event at time $t_0$, i.e., $C(t)=t_0+\beta T_0 (t-t_0)$.
Note that because all active devices are relatively synchronized using the same two synchronization events, small variations in the synchronization event periodicity $T_\text{SYNC}$ do not affect the relative synchronization of the devices. 
\subsubsection{Transmission Phase}
The transmission phase (TX phase)  begins immediately after the synchronization phase terminates, and lasts until the next synchronization event. This is to ensure that all active devices in a TX phase have been relatively synchronized using the last two successive synchronization events. During the transmission phase, the time is divided into $M$ identical timeslots, with timeslot duration designed to contain a message transmission and a guard time. An active device randomly selects a timeslot $i\in\left[0, M-1\right]$ for its transmission attempt at time $C(t_0 + i\cdot \left(\text{ToA}+T_g\right))$. The end of a transmission phase must occur before the arrival of the next synchronization event, marking  the beginning of another transmission phase. 

At any point in time, there can be devices in each of these three phases, while the transition from synchronization to transmission phase is aligned with the reception of a synchronization event. Upon the arrival of a synchronization event, the active devices in the synchronization phase switch to the transmission phase. In contrast, the active devices in the transmission phase enter the sleep phase.

Note that despite the modification to the MAC mechanism of the devices, no LoRaWAN rule or duty-cycle restriction is violated. Accordingly, the proposed approach is transparent to the legacy LoRaWAN backend infrastructure. 

\begin{figure}[!t]
	\begin{algorithm}[H]\small
		\caption{Pseudocode of the Synchronization and Transmission Scheme} 
		\label{alg:ALG1}
		\begin{flushleft}
			\textbf{INPUT:} 
			{$T_\text{SYNC}$: nominal synchronization event periodicity,}

			\hspace*{10mm}{$M$: number of slots in a $T_\text{SYNC}$, \textit{msg}: messages for the GW,}
			
			\hspace*{10mm}{$C(t)$: local clock value at time $t$}
		\end{flushleft}
		
		\begin{flushleft}
			\textbf{Step 1:} Sleep State---waiting for a message for the GW \\
			{\addtolength{\leftskip}{4 mm}
				
			}
		\end{flushleft}
		
		\begin{algorithmic}[1]
			\While { \textit{msg}$=\emptyset$ }
			\State Wait \Comment{\textit{wait for a message to transmit to the GW}}
			\EndWhile 
			\State Jump to \textbf{Step 2} 
		\end{algorithmic}
		\begin{flushleft}
			\textbf{Step 2:} Synchronization---estimation of clock parameters $\left(t_{-1}, t_0, \beta\right)$
			\\{\addtolength{\leftskip}{4 mm}
				
			}
		\end{flushleft}
		
		\begin{algorithmic}[1]
			\State Wait for the 1\textsuperscript{st} synchronization event
			\State $t_{-1}\leftarrow{}C(t)$
			\State Wait for 2\textsuperscript{nd} synchronization event
			\State $t_0\leftarrow{}C(t)$
			\State $\beta\leftarrow{}\frac{t_{0}-t_{-1}}{T_\text{SYNC}}\frac{1}{T_0}$ \Comment{\textit{Estimate local clock rate}}
			\State $C(t)=t_{0}+\beta T_0 (t - t_0)$ \Comment{\textit{Update local clock parameters}}
			\State Jump to \textbf{Step 3} 
		\end{algorithmic}
		
		\begin{flushleft}
			\textbf{Step 3:} Transmission of the message \\
			{\addtolength{\leftskip}{4 mm}
				
			}
		\end{flushleft}
		\begin{algorithmic}[1]
			\State $i\leftarrow{}\text{rand}\left[0,M-1\right]$ \Comment{\textit{Select random timeslot}}
			\State At $C(t_0 + i\cdot \left(\text{ToA}+T_g\right))$ transmit \textit{msg} to GW
			\State Jump to \textbf{Step 1} 
		\end{algorithmic}
	\end{algorithm}
	\vspace{-25pt}
\end{figure}

\section{Throughput Comparison for In-Band and Out-of-Band Synchronization}
\label{sec:th_comparison}
In this section, we derive the theoretical throughput expression of the proposed synchronization and transmission scheme. We consider a network consisting of $N$ EDs and a single GW. In the derivation, we assume an ideal channel and use an ideal collision model without LoRa power-capture.Additionally, to determine the advantage of our scheme, we compare its throughput with that of a generic slotted ALOHA protocol that uses a two-way message exchange for in-band synchronization. 
Later, in Sec.~\ref{sec:model}, the throughput expression is updated to consider the effect of LoRa frame structure on the outcome of collisions and used for the design of the parameters of the scheme.
Note that the assumptions used in this section are only for relative theoretical comparison and parameter design purposes. For the final evaluation of the proposed scheme, in Sec.~\ref{Results}, we consider realistic channel and LoRa capture effects that are commonly used in the literature.

\subsection{Out-of-Band Synchronization}
\label{sec:oob_mod}
Upon generating a message, an ED starts the synchronization process, lasting until two successive synchronization events are received.  
After the synchronization is completed, the ED transmits its message to the GW during the TX phase. The TX phase is itself delimited by two successive synchronization events and its duration is $T_\text{SYNC}$. To respect the activity factor constraint, the mean time between transmissions from the same device must be $\overline{X}\geq {\text{ToA}}/{\alpha}$. The normalized offered traffic in the system $G$, measured in messages per ToA, is given by $G=N\cdot \frac{\text{ToA}}{\overline{X}}$.
The number of active devices $N_A$ during a TX phase of duration $T_\text{SYNC}$ is a Poisson distributed random variable
\begin{equation}
\Psi\left(N_A=k\right)=\frac{\left(G\frac{T_\text{SYNC}}{\text{ToA}}\right)^k e^{-G\frac{T_\text{SYNC}}{\text{ToA}}}}{k!}\text{.}
\label{eq:Psi}
\end{equation}
Each active ED randomly and independently chooses a timeslot in the TX phase for its transmission. Assuming that a TX phase consists of $M$ timeslots, the slot access probability of an active ED is $1/M$. Given $k$ active EDs, the  number of messages sent in one timeslot is a Binomial random variable
\begin{equation}
B_{k,M}\left(i\right)=\binom{k}{i}\left(\frac{1}{M}\right)^i\left(1-\frac{1}{M}\right)^{k-i}\text{.}
\label{eq:B}
\end{equation}
An ED is successful in transmitting its message only if no other active ED choses the same timeslot. Therefore, the normalized average throughput in messages per timeslot is 
\begin{equation}
\label{eq:S'222}
S'=\sum_{k=1}^\infty \Psi\left(k\right) B_{k,M}\left(1\right)\text{.}
\end{equation}
After substituting \eqref{eq:Psi} and \eqref{eq:B}  in \eqref{eq:S'222},  we obtain
\begin{equation}
S'=\frac{G}{M}\frac{T_\text{SYNC}}{\text{ToA}}e^{-G\frac{T_\text{SYNC}}{\text{ToA}}\frac{1}{M}}\text{.}
\label{eq:S'}
\end{equation}
The number of timeslot in a contention period is given by
\begin{equation}
M=\frac{T_\text{SYNC}-\Delta}{\text{ToA}+T_g}\text{,}
\label{eq:M}
\end{equation}
where 
\begin{equation}
\Delta= \max\left(T_\text{SYNC}-\floor*{\frac{T_\text{SYNC}}{\text{ToA}+T_g}}\cdot \left(\text{ToA}+T_g\right) , 2{\delta}\right)\text{, \nonumber
\label{eq:Delta}}
\end{equation} 
is the average portion of a TX phase that is not part of the contention period, ${\delta}$ is the uncertainty in the synchronization event periodicity {experimentally characterized in Sec.~\ref{TimingErrors},} and $\floor*{\cdot}$ is the floor function.
Finally, the normalized channel throughput with out-of-band synchronization, measured in messages per ToA, is given by
{\medmuskip=0mu
	\thinmuskip=0mu
	\thickmuskip=0mu
\begin{equation}
	\begin{aligned}
	S=\frac{\text{ToA}}{\text{ToA}+T_g}\frac{T_\text{SYNC}-\Delta}{T_\text{SYNC}}S'\approx Ge^{-G\frac{T_\text{SYNC}}{T_\text{SYNC}-\Delta}\left(1+\frac{T_g}{\text{ToA}}\right)}\text{.}
	\label{eq:S_oob}
	\end{aligned}
\end{equation}}

\subsection{In-Band Synchronization}
For the analysis of in-band synchronization, we consider the use of a two-way message exchange mechanism as proposed in~\cite{polonelli2019slotted, haxhibeqiri2018low}. Each ED initiates the synchronization procedure by transmitting an uplink synchronization request to the GW. The GW responds with a message (e.g., an acknowledgment) containing  the time at which the uplink transmission was received. This information is used by the ED to synchronize its local clock time to the GW. Like for out-of-band synchronization, the aggregate data traffic generated by the EDs is assumed to have a Poisson distribution. The normalized offered data traffic in the system $G$, measured in messages per ToA, is given by $G=N\cdot \frac{\text{ToA}}{\overline{X}}$. We assume that the synchronization requests generated by the EDs are Poisson distributed.  Considering a mean synchronization interval of $T_\text{SYNC}$ and assuming that both synchronization request and response occupy the channel for a ToA duration, the normalized offered synchronization traffic in the system $G_s$, measured in messages per ToA, is given by $G_s=N\cdot\frac{2\text{ToA}}{T_\text{SYNC}}$.

The aggregate traffic arrival is the result of the transmission of data and synchronization messages. To respect the activity factor constraint, the mean time between data transmission must be $\overline{X}\geq {\text{ToA}}/({\alpha-\frac{\text{ToA}}{T_\text{SYNC}}})$. 
The number of active devices $N_A$ in a timeslot of duration $T_g+\text{ToA}$ is a Poisson distributed random variable
{\medmuskip=0mu\thinmuskip=0mu\thickmuskip=0mu
\begin{equation}
\Psi\left(N_A=k\right)=\frac{\left[\left(G+G_s\right)\left(\frac{T_g+\text{ToA}}{\text{ToA}}\right)\right]^k e^{-\left(G+G_s\right)\frac{T_g+\text{ToA}}{\text{ToA}}}}{k!}\text{.}
\label{eq:Psi2}
\end{equation}}

Then, the normalized channel throughput of slotted ALOHA with the in-band  synchronization mechanism,  measured in messages per ToA, is given by the product of normalized  offered  data  traffic by the probability that there are no collisions in a slot,
{\medmuskip=0mu
	\thinmuskip=0mu
	\thickmuskip=0mu
\begin{equation}
	\begin{aligned}
	S=G\cdot \Psi\left(N_A=0\right)=G\cdot e^{-\left(G+G_s\right)\left(1+\frac{T_g}{\text{ToA}}\right)}\text{.}
	\label{eq:S_ib}
	\end{aligned}
\end{equation}}

\subsection{Throughput Ratio}
We define the throughput ratio $\eta$ as the ratio between the throughput for in-band~\eqref{eq:S_ib} and for out-of-band synchronization~\eqref{eq:S_oob}. Assuming $\Delta = 0$, after a few simple algebraic manipulations, the expression of $\eta$ is given by
\begin{equation}
	\begin{aligned}
	\eta=e^{-N\left[\frac{\left(T_{g,\text{o}}-T_{g,\text{i}}\right)}{\overline{X}}-2\frac{\left(\text{ToA}+T_{g,\text{i}}\right)}{T_\text{SYNC,i}}\right]}
	\label{eq:eta}\text{,}
	\end{aligned}
\end{equation}
where $T_{g,\text{i}}$  and $T_{g,\text{o}}$ are the slot guard time for the in-band and out-of-band synchronization, respectfully.
Considering a local clock skew of $\gamma$, to avoid collisions between adjacent timeslots, we must have $T_g\in\left[0,\gamma \times T_\text{SYNC}\right)$. The guard time should be set to zero following a synchronization event and linearly increased until it reaches its maximum value of $\gamma \times T_\text{SYNC}$ immediately before a new synchronization event.
\begin{figure}%
	\centering
	\includegraphics[width=0.9\linewidth]{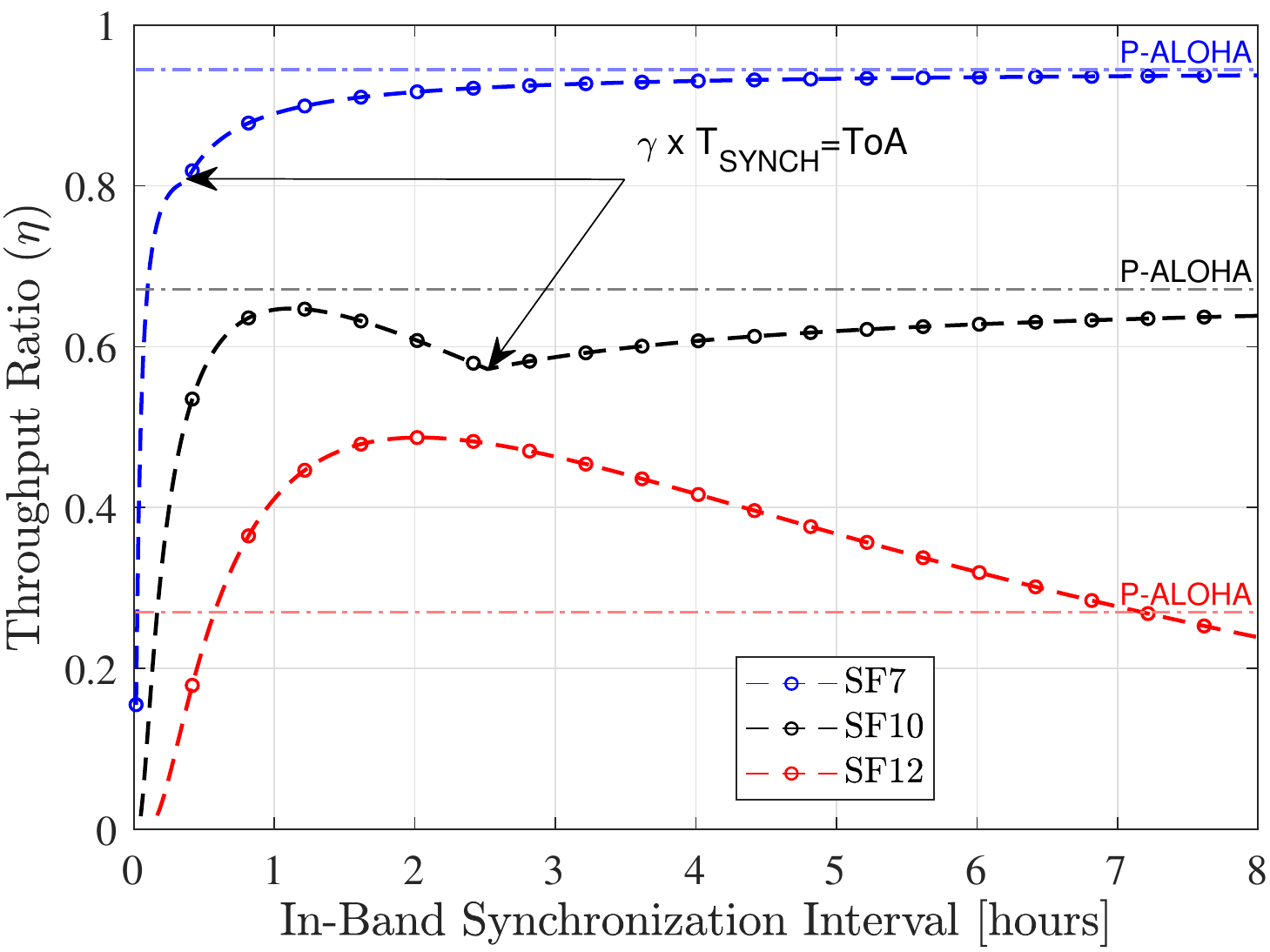}\hfill
	\caption{Throughput ratio; $N=1000$, $\alpha=0.0033$, $\overline{X}=15$ minutes, $\gamma=40$ ppm, $ T_\text{SYNC,o}=60$ seconds.}
	\label{fig:Tsync}
	\vspace{-15pt}
\end{figure}

Fig.~\ref{fig:Tsync} shows the throughput ratio \eqref{eq:eta}, for the parameters given in the caption. It is observed that the throughput obtained using out-of-band synchronization is higher than that obtained by in-band synchronization, i.e., $\eta\leq 1$ for any value of the in-band synchronization interval $T_\text{SYNC,i}$. When the in-band synchronization interval is short, the duty-cycle constraints prevent some of the data  to be transmitted. For long in-band synchronization intervals, the in-band synchronization is ineffective, and the system performs as in pure ALOHA. In general, small SFs, with their shorter ToA, require more frequent synchronization, which affects the channel air-time available to data traffic. In Fig.~\ref{fig:Tsync}, for SF~7, we can observe that under duty-cycle limitations, pure ALOHA channel access is sometimes to be preferred over slotted ALOHA with in-band synchronization. These initial results support the use of out-of-band synchronization to improve LoRa throughput (consequently, the directly related success probability) via synchronous communication. 
The design of the parameters for the transmission phase of the proposed communication scheme, namely the number of timeslots ($M$), the timeslot guard time ($T_g$) and the phase guard time ($\Delta$), is discussed in the next section. 

\section{Co-Design of the Synchronization and  Communication Parameters}
\label{sec:model}
This section presents the design of the parameters for the proposed synchronization and communication scheme based on the characteristics of LoRa frame collisions. To this end, we introduce the probability of inter-slot collision in the analytical model \eqref{eq:S_oob}. In this model's context, a message is successfully delivered to the GW if there are no intra and inter-slot collisions. Intra-slot collisions ($C$) occur if more than one EDs choose the same slot for transmission. Inter-slot collisions occur when messages transmitted in adjacent timeslots partially overlap due to synchronization errors and transmission time uncertainties.
{Considering the low data rate of LoRa and the equivalent long ToA, we can safely assume that realistic values of the synchronization error can cause inter-slot collisions only between adjacent timeslots.} When the reference message is affected by a collision with a message transmitted in the previous timeslot or successive timeslot, we respectively indicate it as a left ($L$) or right ($R$) collision.

To characterize the collisions of LoRa messages, we have to consider the structure of the frame transmitted by LoRa EDs.
Fig.~\ref{fig:frame}.c shows the structure of an uplink LoRa frame. The head and the tail sections of the frame are most likely to be affected by inter-frame collisions. A LoRa frame begins with a preamble, composed of a fixed part and a configurable part of variable length. The fixed part consists of a 2-symbol synchronization word and an additional 2.25 symbols, whereas the configurable part can be adjusted in length between 6 and 65535 symbols. The frame terminates with a 2-byte CRC field for detecting errors in the payload.
To understand the outcome of collisions between two LoRa frames (i.e., Frame~A and Frame~B in  Fig.~\ref{fig:frame}b), experimental studies (e.g., \cite{rahmadhani2018lorawan, haxhibeqiri2017lora, bor2016lora})  concluded that: 1) a frame can be successfully decoded if it arrives one ToA before and up to at most $\text{ToA}-\left(T_p-5T_\text{sym}\right)$ after the other frame, with $T_p$ the preamble duration and  $T_\text{sym}$ the symbol time, and 2) for overlaps longer than $T_p-5T_\text{sym}$, no transmission is correctly received unless it is for the power-capture effect. Therefore, the receiver requires a minimum of 5 symbols to detect the preamble and synchronize to the transmission. These observations are summarized in Table \ref{tab:collision_table}, where $T_A$ and $T_B$ are the arrival time of the first and second frame, respectively.

\begin{table}[t]
	\caption{Outcome of overlapping LoRa frames of similar power.}
	\centering\setlength\tabcolsep{2.8pt}
	\scalebox{1}{\begin{tabular}{|c|c|c|}
			\cline{1-3}
			\textbf{Condition}  & \textbf{Frame A} & \textbf{Frame B} \\\hline
			\multicolumn{1}{|l|}{$T_A+\text{ToA}<T_B$}&    Successful     &    Successful     \\\hline
			\multicolumn{1}{|l|}{$T_B\leq T_A+\text{ToA}< T_B+\left(T_p-5T_\text{sym}\right)$}&  CRC Error & Successful   \\\hline
			\multicolumn{1}{|l|}{$T_B+\left(T_p-5T_\text{sym}\right)\leq T_A+\text{ToA}$}&  CRC Error  &  Frame dropped\\ \hline     
	\end{tabular}}
	\label{tab:collision_table}
	\vspace{-10pt}
\end{table}

 Based on the observations presented in Table \ref{tab:collision_table}, the model of Sec.~\ref{sec:oob_mod} is modified to consider the probabilities of left and right inter-slot collision. 
In the following, $\text{Pr}(A)$ and $\text{Pr}(\overline{A})$ denote the probability of an event $A$ and its complement, respectively.
Because the number of EDs is finite, the probability of intra- and inter-slot collisions $\text{Pr}\left(C\right)$, $\text{Pr}\left(L\right)$  and $\text{Pr}\left(R\right)$  are dependent.

In a contention period of the proposed synchronization and communication mechanism with $M$ timeslots and $k$ active EDs, the probabilities that a message is not affected by left and right inter-slot collisions, conditioned that there is no intra-slot collision, are given by
\begin{equation}
\begin{aligned}
\text{Pr}\left(\overline{L}\mid \overline{C}\right)&\approx\sum_{j=0}^{k-1}B_{k-1,M-1}\left(j\right)\left(1-p_L\right)^j\text{,}\\
\text{Pr}\left(\overline{R}\mid \overline{C}\right)&\approx\sum_{j=0}^{k-1}B_{k-1,M-1}\left(j\right)\left(1-p_R\right)^j\text{,}
\label{eq:PL}
\end{aligned}
\end{equation}
where $p_L$ and $p_R$ are the probabilities that the synchronization and transmission uncertainties cause a left or right inter-slot collision.
The approximation in \eqref{eq:PL} originates from the statistical dependency of $\text{Pr}\left(\overline{L}\mid \overline{C}\right)$ and $\text{Pr}\left(\overline{R}\mid \overline{C}\right)$.
The normalized average throughput in messages per timeslot is 
\begin{equation}
S'=\sum_{k=1}^\infty \Psi\left(k\right) B_{k,M}\left(1\right)\text{Pr}\left(\overline{L}\mid \overline{C}\right)\text{Pr}\left(\overline{R}\mid \overline{C}\right)\text{,}
\label{eq:eq11}
\end{equation}
{where $B_{k,M}\left(1\right)=P\left(\overline{C}\right)$ is the probability of having only one message transmitted in  a timeslot.}
By using \eqref{eq:PL} and performing a few simple algebraic manipulations, \eqref{eq:eq11} becomes
\begin{equation}
S'=\frac{G}{M}\frac{T_\text{SYNC}}{\text{ToA}}e^{-G\frac{T_\text{SYNC}}{\text{ToA}}\frac{\left(1+p_l\right)\left(M-1\right)+p_R\left(M-2\right)}{M\left(M-1\right)}}\text{,}
\label{eq:S'1}
\end{equation}
where $M$, the number of timeslot in a contention period, is given by \eqref{eq:M}.
{As a matter of fact, it is feasible to assume that $M\gg 1$, since in real scenarios $T_g+\text{ToA}\ll T_\text{SYNC}$, being $T_g < \gamma \times T_\text{SYNC}$.} Using this observation,  \eqref{eq:S'1} can be approximated as
{\medmuskip=-1mu
	\thinmuskip=-1mu
	\thickmuskip=-1mu\begin{equation}
\begin{aligned}
S'&\approx \frac{GT_\text{SYNC}}{T_\text{SYNC}-\Delta}\left(1+\frac{T_g}{\text{ToA}}\right)e^{-\frac{GT_\text{SYNC}}{T_\text{SYNC}-\Delta}\left(1+\frac{T_g}{\text{ToA}}\right)\left(1+p_L+p_R\right)}\text{.}
\end{aligned}
\end{equation}}
Finally, the normalized channel throughput measured in message per ToA is given by
\begin{equation}
	\begin{aligned}
	S= Ge^{-G\frac{T_\text{SYNC}}{T_\text{SYNC}-\Delta}\left(1+\frac{T_g}{\text{ToA}}\right)\left(1+p_L+p_R\right)}\text{,}
	\label{eq:S}
	\end{aligned}
\end{equation}
where only $p_L$ and $p_R$ remain to be found.

\subsection{Inter-Frame Collision Probability and Optimal Guard Time}\label{SLoRamodel}
In this section, we find $p_L$ and $p_R$, the left and right inter-slot collision probabilities for the proposed LoRa time-slotted communication. As previously discussed, inter-slot collisions are typically caused by the uncertainties in the synchronization and transmission times. 
Time uncertainty can be described by its probability distribution; in particular, we  assume two likely distributions of the timing error: Gaussian  and uniform distribution.
\begin{figure*}%
	\centering
	\subfloat[Spreading factor 7 {($T_\text{SYNC}=1$ min)} ]{\includegraphics[width=0.33\linewidth]{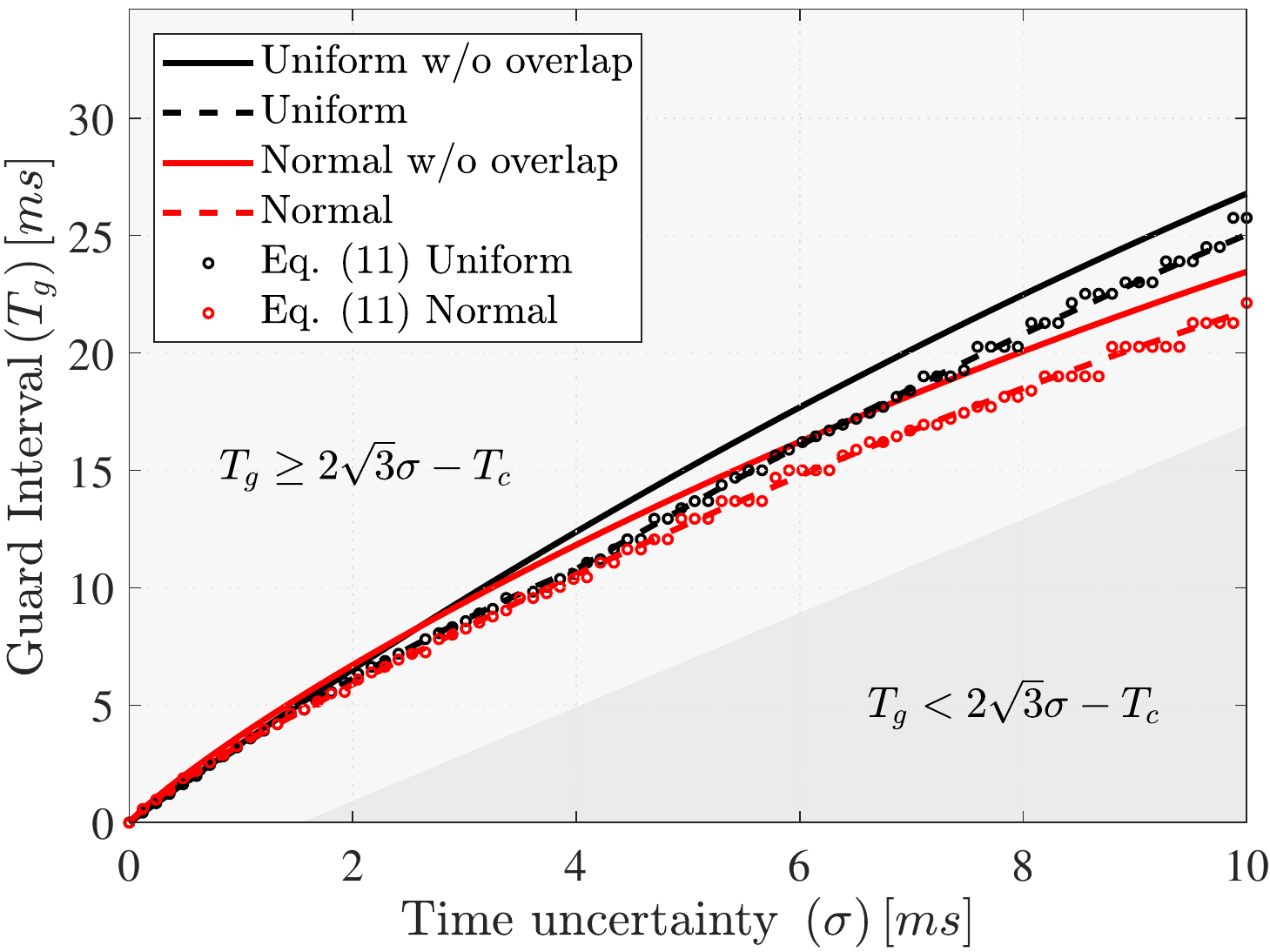}\label{fig:optimal_solution_a}}\hfill
	\subfloat[Spreading factor 10 {($T_\text{SYNC}=5$ min)}]{\includegraphics[width=0.33\linewidth]{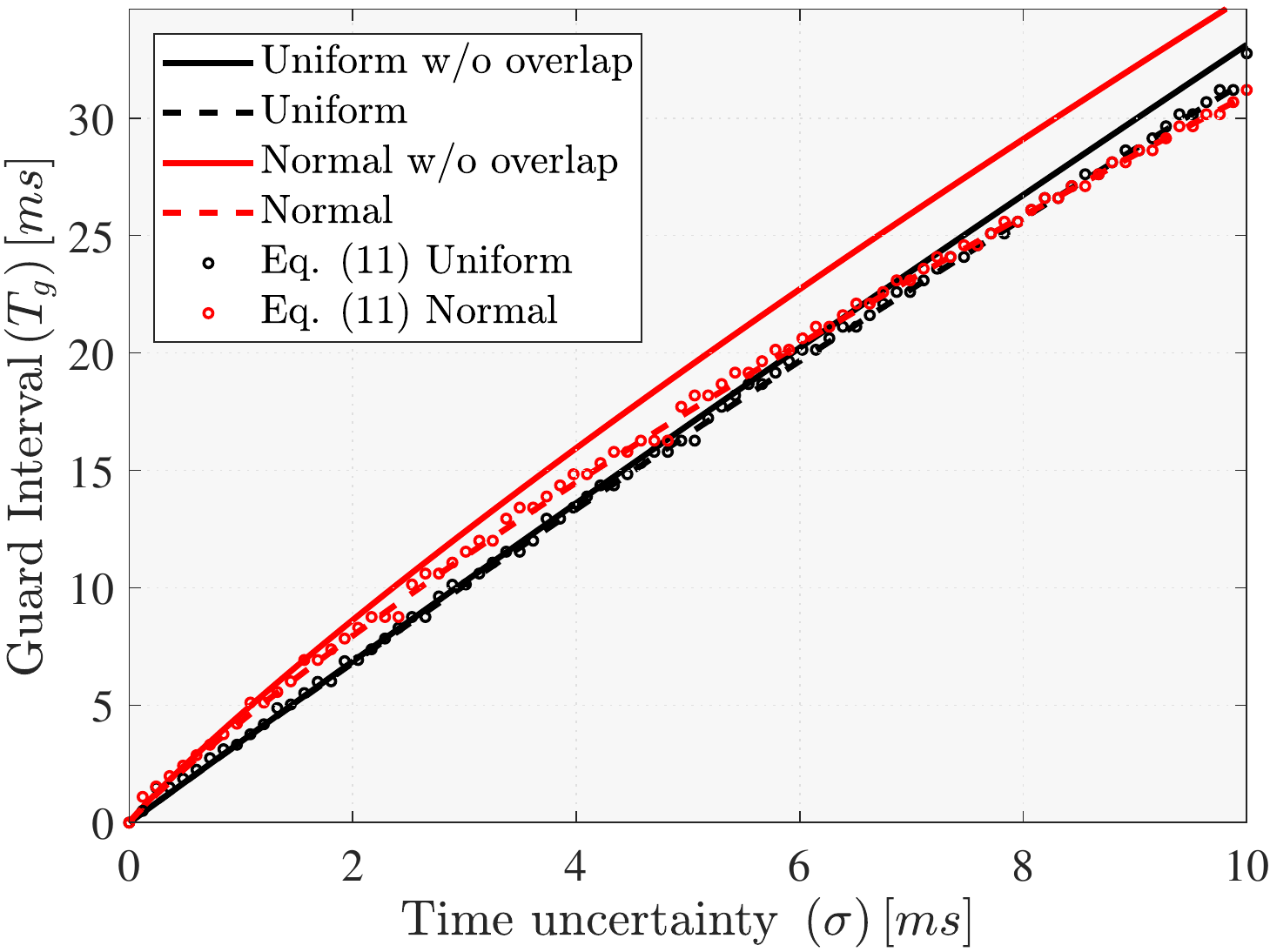}\label{fig:optimal_solution_b}}\hfill
	\subfloat[Spreading factor 12 {($T_\text{SYNC}=10$ min)}]{\includegraphics[width=0.33\linewidth]{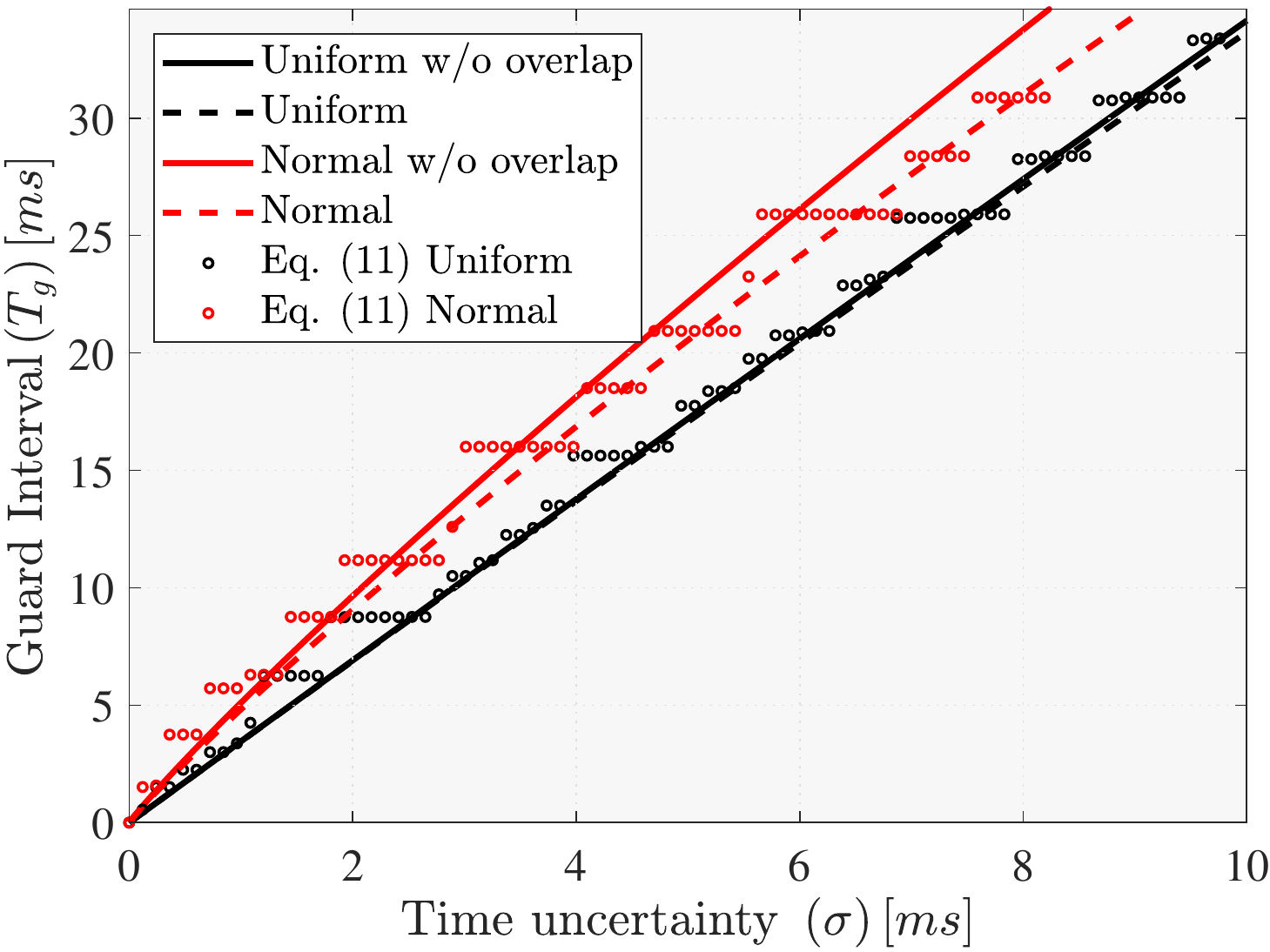}\label{fig:optimal_solution_c}}
	\caption{Optimal guard interval $T_g^*$, for uniform and Gaussian distributed timing errors with standard deviation $\sigma$. The guard time without overlap is to be interpreted as the value obtained by considering LoRa frames with a preamble of 5 symbols (i.e. $T_c$=0). {With the exception of $T_\text{SYNC}$, the parameters used to obtain the results are reported in Table~\ref{tb:par}.} }
	\label{fig:optimal_solution}
	\vspace{-15pt}
\end{figure*}
\subsubsection{Gaussian Distributed Timing Error}
The time uncertainty affecting the transmission of a message is the result of the cumulative effects of the timing errors. Consequently, by the central limit theorem, a Gaussian distribution is a natural choice for the probability distribution of the uncertainty.
{We assume a Gaussian distributed time uncertainty  with zero mean and standard deviation $\sigma$ (i.e., $\mathcal{N}\left(0,\sigma^2\right)$).} The timing errors of two messages occupying adjacent slots are independent, and the bivariate distribution of the time uncertainty of the two messages is 
\begin{equation}
f_{X,Y}\left(x,y\right)=
\frac{1}{2\pi \sigma^2}e^{-\frac{x^2+y^2}{2{\sigma^2}}} \text{.}
\end{equation}
Based on~\cite{crozier1990sloppy}, the probability of inter-slot collision is
\begin{equation}\label{eq:p_tau}
\Phi\left(\tau\right)=\text{Pr}\left(X>Y+\tau\right)=Q\left(\frac{\tau}{\sqrt{2}\sigma}\right)\text{,}
\end{equation}
where $Q(\cdot)$ is the Gaussian $Q$-function, and $\tau$ is the value of timing error for which inter-slot collisions start to occur. The amount of timing error necessary to cause a collision is different depending on which part of the LoRa frame is affected. Overlaps with a message transmitted in the successive (right)  timeslot affect the tail of the transmitted frame containing the CRC field and causing a CRC error.  Consequently,  timing errors larger than the timeslot guard time ($T_g$) cause a CRC error, which is interpret as a failed transmission. The probability of a right collision can be found from \eqref{eq:p_tau} as
\begin{equation}
p_R=\Phi\left(T_g\right)=Q\left(\frac{T_g}{\sqrt{2}\sigma}\right)\text{.}
\end{equation}
Overlaps  with a message transmitted in the previous (left)  timeslot affect the preamble of the transmitted frame, preventing the demodulator from synchronizing to the preamble.  As reported in Table~\ref{tab:collision_table}, the message is not correctly received if the overlap is greater than $T_c=T_p-5T_\text{sym}$.
The probability of a left collision can be found from \eqref{eq:p_tau} as 
\begin{equation}
p_L=\Phi\left(T_g+T_c\right)=Q\left(\frac{T_g+T_c}{\sqrt{2}\sigma}\right)\text{.}
\end{equation}

\subsubsection{Uniform Distributed Timing Error}
In alternative to the Gaussian distributed timing error, we also investigate a uniformly distributed error with zero mean and standard deviation $\sigma$ (i.e. $\mathcal{U}\left(-\sqrt{3}\sigma,+\sqrt{3}\sigma\right)$).
Assuming independent timing error in adjacent messages, the bivariate distribution of the time uncertainty is 
\begin{align}
f_{X,Y}\left(x,y\right)=&
\frac{1}{\left(2{\sqrt{3}\sigma}\right)^2} & \text{for } & -{\sqrt{3}\sigma }\leq x,y\leq {\sqrt{3}\sigma }\text{.}
\end{align}
For a guard time $T_g$, the inter-slot collision probability can be expressed as
\begin{equation}
\Phi\left(\tau\right)=\text{Pr}\left(X>Y+\tau\right)=\frac{\left[\max\left\{0,2{\sqrt{3}\sigma}-\tau\right\}\right]^2}{2\left(2{\sqrt{3}\sigma}\right)^2}\text{.}
\end{equation}
Based on the observations in Table~\ref{tab:collision_table}, the probability of right and left inter-slot collision can be found as 
\begin{equation}
\begin{aligned}
p_R&=\Phi\left(T_g\right)=\frac{\left[\max\left\{0,2{\sqrt{3}\sigma}-T_g\right\}\right]^2}{2\left(2{\sqrt{3}\sigma}\right)^2}\text{,}\\
p_L&=\Phi\left(T_g+T_c\right)=\frac{\left[\max\left\{0,2{\sqrt{3}\sigma}-\left(T_g+T_c\right)\right\}\right]^2}{2\left(2{\sqrt{3}\sigma}\right)^2}\text{.}
\end{aligned}
\label{eq:p_Rp_L}
\end{equation}

\subsubsection{Optimal Guard Time}
{Under the aforementioned assumption that $M\gg 1$, it is possible to obtain the optimal duration of the guard interval that maximizes the normalized channel throughput, by combining \eqref{eq:S} and \eqref{eq:p_Rp_L}.}
For uniformly distributed timing error, the optimal guard time $T_g^*$ is given by solving
{\medmuskip=0mu
\thinmuskip=-2mu
\thickmuskip=-2mu
\nulldelimiterspace=0pt
\scriptspace=0pt
\begin{equation}
	\argmin_{T_g} \left(\mkern-6mu1+\frac{T_g}{\text{ToA}}\mkern-4mu\right)\mkern-5mu\left(\mkern-6mu1+\frac{\left[2{\sqrt{3}\sigma}-\left(T_g+T_c\right)\right]^2\mkern-6mu+\left[2{\sqrt{3}\sigma}-T_g\right]^2}{2\left(2{\sqrt{3}\sigma}\right)^2}\right)\text{.}
	\label{eq:Tg_opt}
\end{equation}}\noindent
In the case of Gaussian distributed timing error,  the throughput function is twice differentiable  and concave with respect to $T_g$ and the optimal guard time can be found by numerically solving
{\medmuskip=0mu
	\thinmuskip=0mu
	\thickmuskip=0mu
	\begin{equation}
	\begin{aligned}
	\frac{\left(\text{ToA}+T_g^*\right)}{2\sigma\sqrt{\pi}}&\left({\text{exp}\left(-\frac{T_g^{*^2}}{4\sigma^2}\right)+\text{exp}\left(-\frac{\left(T_g^*+T_c\right)^2}{4\sigma^2}\right)}\right)\\&-\frac{1}{2}\text{Erfc}\left(\frac{T_g^*}{2\sigma}\right)-\frac{1}{2}\text{Erfc}\left(\frac{T_g^*+T_c}{2\sigma}\right)=1\text{.}
	\end{aligned}
	\label{eq:Tg_opt_n}
	\end{equation}
}
The optimal guard time in the case of Gaussian and uniformly distributed timing error  with standard deviation $\sigma$, derived respectively from \eqref{eq:Tg_opt} and \eqref{eq:Tg_opt_n} are shown in Fig.~\ref{fig:optimal_solution} for SF 7, 10 and 12{, and increasingly long synchronization intervals}. {The numerically obtained optimal $T_g^*$ from the exact throughput expression in \eqref{eq:eq11} closely matches the solution of \eqref{eq:Tg_opt} and \eqref{eq:Tg_opt_n} for SF~7 and SF~10. However, due to long ToA of higher SFs,  the approximations \eqref{eq:Tg_opt} and \eqref{eq:Tg_opt_n} become tight only by increasing $T_\text{SYNC}$ to satisfy the condition $M\gg 1$.} We can observe that for the same time uncertainty, larger SFs require longer guard time because of their longer ToA. Considering the probability that a message survives partial preamble collisions allows to reduce the guard time especially for smaller SFs. {In Fig.~\ref{fig:optimal_solution_b} and \ref{fig:optimal_solution_c}, a longer guard interval is required for Gaussian distributed timing errors, conversely in Fig.~\ref{fig:optimal_solution_a} the reverse is observed. This is consequence of the fact that the increased overhead resulting from a shorter ToA makes the use of longer guard intervals unfeasible for reducing the inter-slot collision caused by the tail of the Gaussian distribution.} 

In the next section, we estimate the time uncertainty $\sigma$, considering all possible sources of error that can affect the transmission of a message in the proposed synchronous communication.

\section{{Timing Errors Sources}}\label{TimingErrors}
For any synchronization mechanism, timing errors arise from the transceivers' non-ideal behavior, the propagation delay of the messages, and the actual synchronization mechanism. {In the following, the contributions to the uncertainty of an ED transmission time are discussed and characterized using Type~A (i.e., repeated measurements) and Type~B uncertainty analysis \footnote{{Type~A uncertainty is estimated by a statistical analysis of measured quantity (i.e., repeated measurements). Type~B uncertainty instead is estimated based on all of the available data collected from anything other than an experiment\cite{jcgm:2008:EMDG}.}}}

\subsection{Transmission Timing Error}
A LoRa transceiver is responsible for transmitting the LoRa frame once the timeslot boundary has been reached. Unfortunately, traveling across the lower levels of the device protocol stack adds additional delays that cannot be entirely removed even through proper calibration. Indeed, the delay depends on the adopted software/hardware. For this reason, the Type~A uncertainty $u_{TX}$ must be taken into account.  In this paper, we considered a Semtech SX1272 transceiver, for which the transmission uncertainty is in the order of few microseconds irrespective of the SF~\cite{rizzi2017evaluation}. 
\subsection{Frame Propagation Delay}
Two frames transmitted precisely  at the start of their respective timeslots can still collide at the receiver due to the different propagation delay. 
In the case of line-of-sight (LOS) condition, the propagation delay increases linearly with the distance of a device from the GW. The propagation delay of an ED at a distance $r$ from the GW is $r/v$, where $v$ is the speed of light. Assuming that the EDs are uniformly distributed in an annular shaped region, with radii $R_L$ and $R_l$, the distribution $f_{PD}(x)$ of the propagation delay for an ED is given by 
\begin{align}\label{fpd_x}
f_{PD}\left(x\right) &=
\frac{2xv^2}{R_L^2-R_l^2} & \text{for } &\frac{R_l}{v}\leq x\leq \frac{R_L}{v}\text{.}
\end{align}
If the EDs are distributed on a disk-shaped region with radius $R$, we have $R_L=R$ and $R_l=0$, resulting in a propagation delay of mean $\mu_{PD}=\frac{2}{3}\frac{R}{v}$ and standard deviation $\sigma_{PD}=\frac{R}{3\sqrt{2}v}$, respectively.
The Type~B uncertainty $u_{PD}$ can be estimated as 
\begin{equation}\label{eq:u_PD}
u_{PD}=\sqrt{\mu_{PD}^2+\sigma_{PD}^2},
\end{equation}
the average value is taken into account since it is not known apriori. Thus, when $R$ is in the order of a few kilometers, $u_{PD}$ is in the order of few microseconds.
{To account for the non-LOS and multipath conditions, common in urban and sub-urban scenarios, one possibility is to update \eqref{eq:u_PD} based on the RMS delay spread~\cite{7c51be827a654249a0fdbd26fcd4c47a}.}

\subsection{Local Clock Error}
\label{sec:clk_error}
From a device's perspective, the start of the next transmission at $t_d$ in the true time reference is seen as the local clock value $C(t_d)$. If a synchronization procedure, as the one described in Sec.~\ref{sec:Sync_alg}, is carried out by the device, the beginning of the TX phase correspond with the detection of the synchronization event at time $t_0$. The clock value at time $t_d$, relative to the beginning of the current TX phase, is given by
\begin{equation}\label{actualct}
V(t_d)=C(t_d)-C(t_0)=\beta (t_d-t_0)T_0\text{,}
\end{equation}
where $t_0$ is the true clock value at the origin of the current synchronization interval, $\beta$ is the clock rate, $T_0$ is the nominal clock period, and a first-order $C(t)$ model is adopted (SKM model~\cite{SKM}).

Thus, the timing error affecting a device relative to the start of a frame transmission is the result of the three different contributing factors: the uncertainty in the syntonization procedure ($u_\beta = \frac{\beta}{T_\text{SYNC}}\cdot u_{T_\text{SYNC}}$) causing the transmitter to drift from the ideal rate, the uncertainty $u_{t_0}$ in the $t_0$ instant (e.g., due to the detection mechanism of the synchronization event, the propagation delay of the signal on air), and the uncertainty $u_{t_d}$ in the transmission time $t_d$.
The overall uncertainty $u_V$ can be obtained using propagation of uncertainty as
\begin{equation}\label{overalluncertainty}
u_V=\sqrt{\left(\frac{\partial {V}}{\partial{\beta}}\right)^2\!\!\cdot u_\beta^2  + \left(\frac{\partial {V}}{\partial{t_d}}\right)^2\!\!\cdot u_{t_d}^2 + \left(\frac{\partial {V}}{\partial{t_0}}\right)^2\!\!\cdot u_{t_0}^2}\text{.}
\end{equation}
Assuming the worst case scenario, i.e., $t_d-t_0=T_\text{SYNC}$, the overall uncertainty $u_V$ can be obtained from~\eqref{overalluncertainty} as 
{\medmuskip=-1mu
	\thinmuskip=0mu
	\thickmuskip=0mu
	\begin{equation}\label{uncertainty}
	\begin{aligned}
		u_V&=T_0\sqrt{\left(t_d-t_0\right)^2\cdot \left(\frac{\beta}{T_\text{SYNC}}\right)^2 \cdot u_{T_\text{SYNC}}^2  + \beta ^2\left(u_{t_d}^2 + u_{t_0}^2\right)}\\
		&=\beta T_0\sqrt{u_{T_\text{SYNC}}^2 + u_{t_d}^2 + u_{t_0}^2}\text{.}
	\end{aligned}
	\end{equation}}
\begin{figure}%
	\centering
	\includegraphics[width=0.9\linewidth]{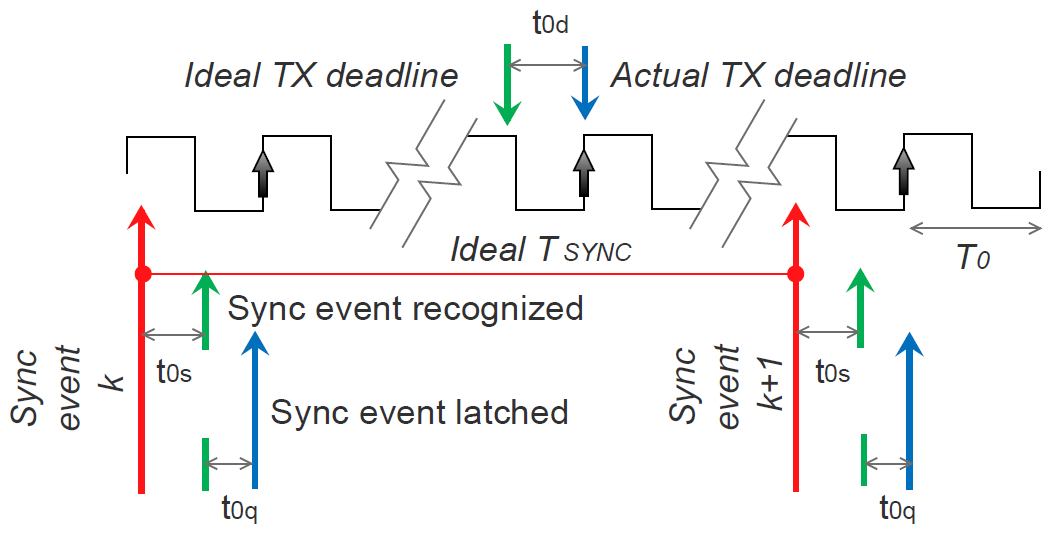}\hfill
	\caption{A sketch of the timing errors affecting the detection of an out-of-band synchronization event.  $t_{0s}$ is the error introduced by the detection of the synchronization event at the out-of-band receiver, $t_{0q}$ is the error introduced by the local clock sampling.}
	\label{fig:timing_errors}
	\vspace{-15pt}
\end{figure}
\begin{figure*}[!t]%
	\centering
	\subfloat[Spreading factor 7 ]{\includegraphics[width=0.332\linewidth]{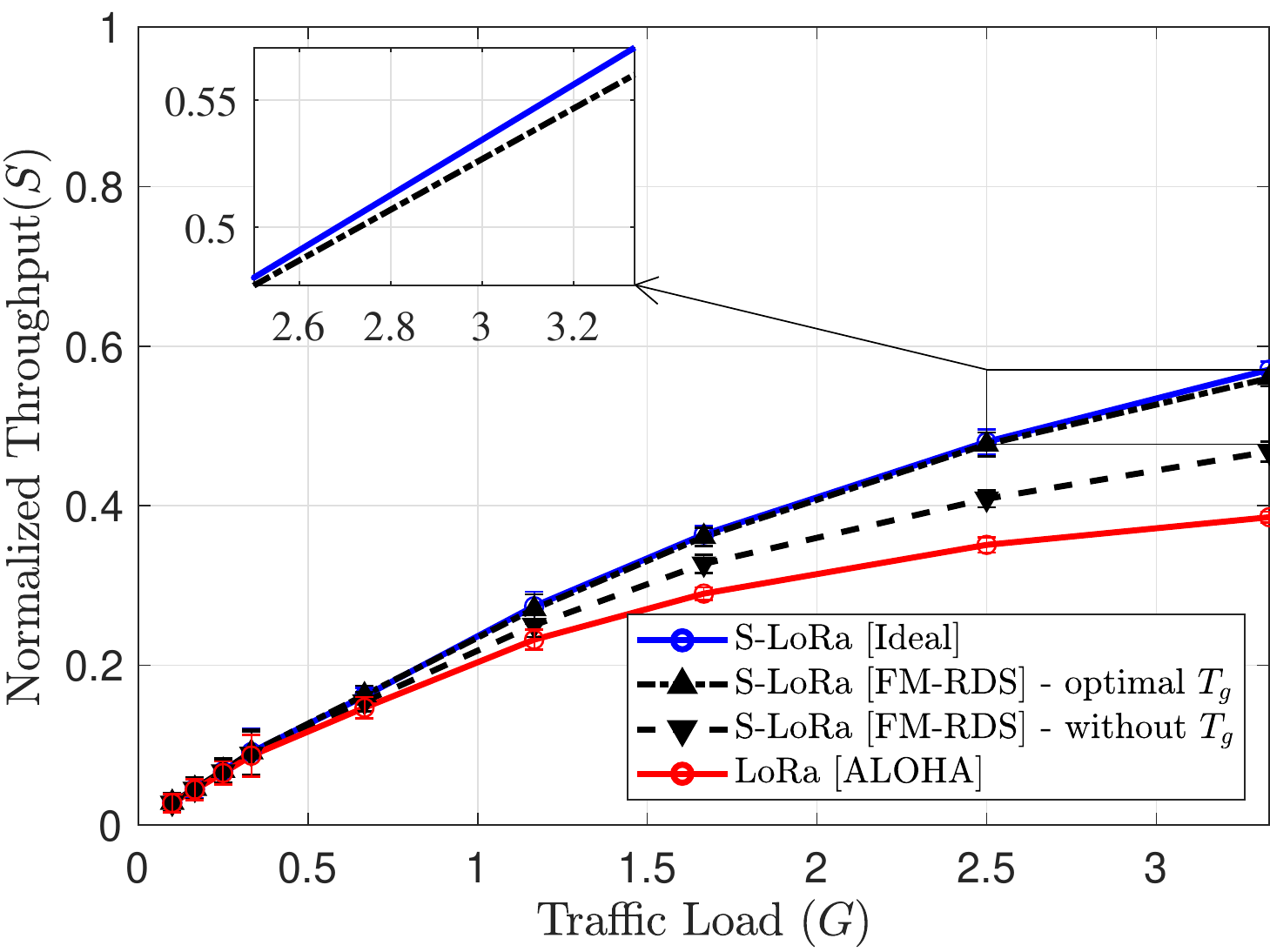}}\hfill
	\subfloat[Spreading factor 10]{\includegraphics[width=0.332\linewidth]{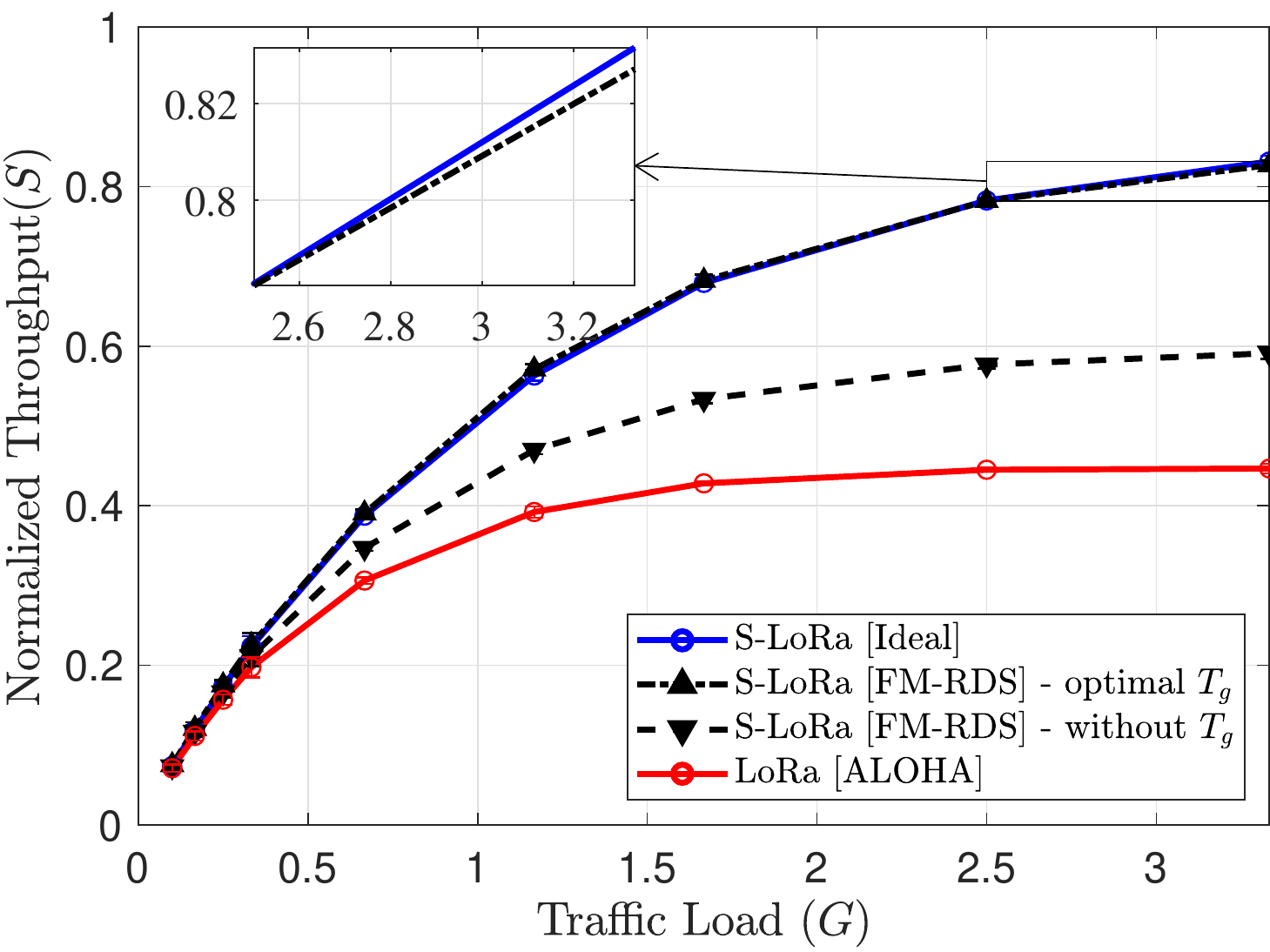}}\hfill
	\subfloat[Spreading factor 12]{\includegraphics[width=0.332\linewidth]{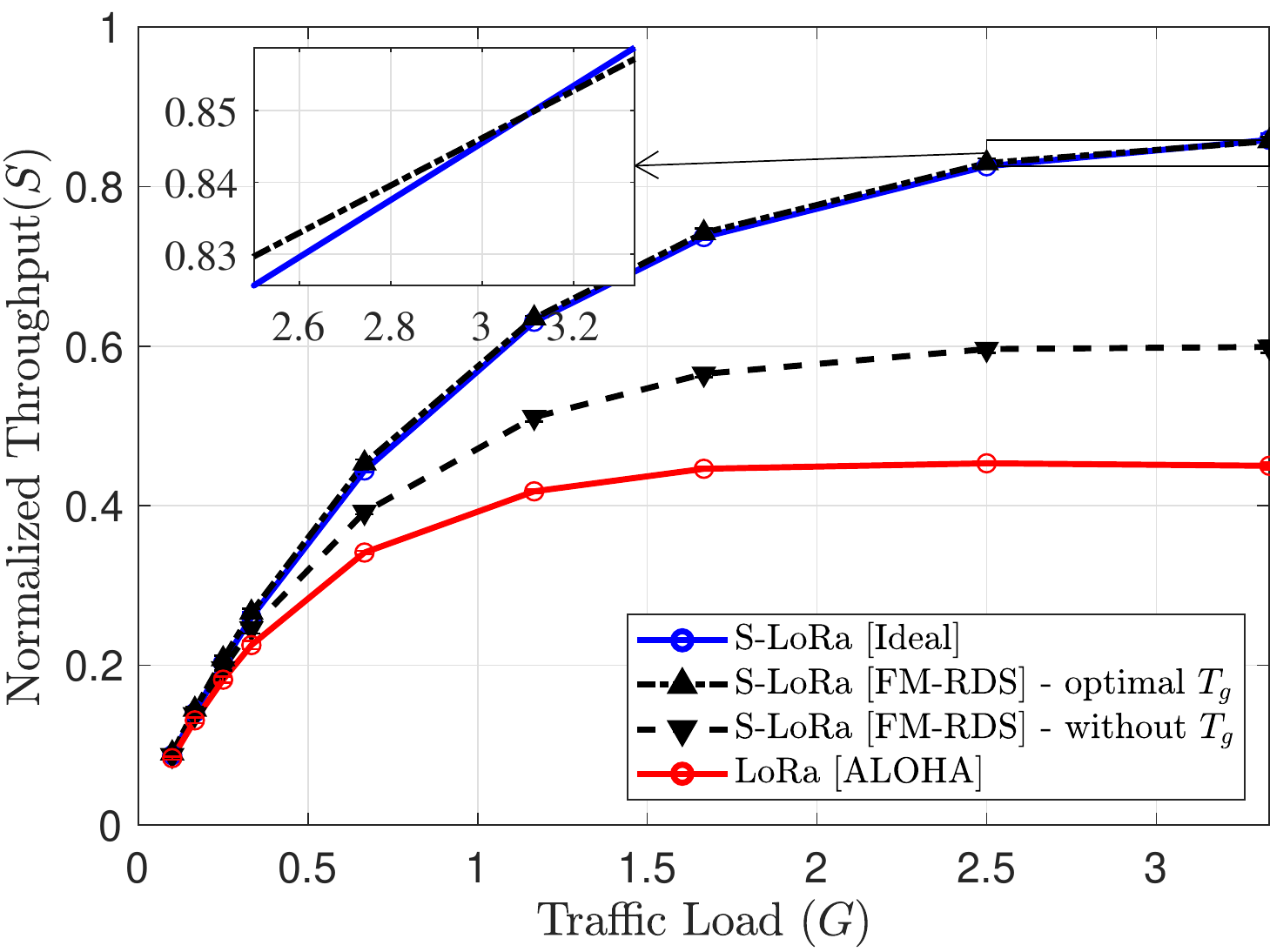}}
	\caption{LoRa network normalized  channel  throughput  for the asynchronous ALOHA communication (LoRA~[ALOHA]) and for the proposed synchronous communication  (S-LoRa). {For the considered duty-cycle, the traffic load ($G$) is generated by a number of devices $N= \left\lbrace 30,  50, 75, 100, 200, 350, 500, 750, 1000\right\rbrace$.}  S-LoRa~[Ideal] represents the case without synchronization errors whereas  S-LoRa~[FM-RDS] is the case with FM-RDS synchronization errors.}
	\label{fig:th}
	\vspace{-10pt}
\end{figure*}

It is possible to further elaborate the uncertainty contributions considering the diagram in Fig. \ref{fig:timing_errors}, where it is assumed that the device is able to latch the synchronization events and to transmit a new frame only on the rising edge of the local clock. Thus, errors $t_{0s}$, $t_{0q}$ and $t_{0d}$ arise. We can assume that both $t_{0q}$ and $t_{0d}$ are $\mathcal{U}[0,T_0]$ and the value of $T_0$ is generally known. On the contrary, the quantity $u_{t_0s}$ must be evaluated, e.g., by performing experiments in agreement with Type~A uncertainty estimation. Consequently, $u_{T_\text{SYNC}}=\sqrt{4u_{t_0q}^2+4u_{t_0s}^2}=\sqrt{4\frac{T_0}{3}^2+4u_{t_0s}^2}$, $u_{t_0}=\sqrt{u_{t_0q}^2+u_{t_0s}^2}=\sqrt{\frac{T_0}{3}^2+u_{t_0s}^2}$, and $u_{t_d}=T_0/\sqrt{3}$. 

At this point, we consider using the periodic CT-group transmitted by \mbox{FM-RDS} for synchronizing the devices.  
We experimentally characterize the uncertainty in the reception of the synchronization event $u_{t_0s}$ by measuring, using a GPS-synchronized time server, the offset in time between the interrupt generated by the same CT-group at two Silicon Labs Si4703 \mbox{FM-RDS} decoders. We generally observed that $T_0\ll u_{t_0s}$, as a result,  \eqref{uncertainty} can be simplified to
\begin{equation}\label{uvworstsimplify}
\begin{aligned}
u_V\approx\sqrt{5}\cdot u_{t_0s}.
\end{aligned}
\end{equation}

By tuning the two decoders on the Italian broadcasting station "RTL102.5" an uncertainty of $u_{t_0s}\approx 0.34$ ms was measured.
From the same evaluation,  we found that the uncertainty in the CT-group transmission periodicity, which affects the design of $\Delta$ in ~\eqref{eq:M} was ${\delta} \approx 200$~ms.

\subsection{Concluding Remarks}
All the previously addressed errors contribute to the overall timing error introduced in Sec.~\ref{SLoRamodel}. Without loss of generality, they can be grouped into the overall time uncertainty as
\begin{equation}\label{u}
u=\sqrt{u_{TX}^2 + u_{PD}^2 + u_{V}^2}.
\end{equation}

The $u$ value can then be transformed into an equivalent standard deviation $\sigma$ of a probability density function---describing the possible timing error population. We used it in the design of the guard time according to Sec.\ref{sec:model}, and for simulating the slotted communication under time uncertainty. 
It has to be emphasized that the proposed approach relies on the assumption of simultaneous estimation of the local clock rate by all the devices, so that $T_\text{SYNC}$ variations affect all of them, without compromising the timeslot borders.

\section{Communication Performance Evaluation: Simulation Results}
\label{Results}
In this section, we report the performance metrics of the proposed transmission mechanism obtained from simulations. To this end, we modified LoRaSim~\cite{bor2016lora} simulator to incorporate our proposed communication scheme. 
Table~\ref{tb:par} lists the typical parameters, which were used in the simulation~\cite{9018210,7803607}.
Without changing the frame collision and capture model of LoRaSim, we model the wireless channel using Rayleigh fading and log-distance path-loss with exponent of three to represent a sub-urban scenario. {Note that both intra and inter-slot collisions have been implemented into the collision model of LoRaSim.}
We model the time uncertainty affecting each transmission as an independent and identically distributed  stochastic offset; the  standard deviation of the time offset distribution is the worst-case compound uncertainty ($\sigma$).
The  modeling of the timing errors using the uncertainty obtained from Sec.~\ref{TimingErrors}, allowed us to simulate scenarios with a high number of devices while avoiding the computationally expensive task of modeling the local clock of each device. 

\begin{table}[!htb]
	\caption{LoRa Simulations Parameters}
	\centering
	\scalebox{0.87}{\setlength\tabcolsep{1.7 pt}
		\begin{tabular}{lllllll}
			\cline{1-3} \cline{5-7}
			\multicolumn{1}{|l|}{\textbf{Parameter}} & \multicolumn{1}{l|}{\textbf{Sym.}} & \multicolumn{1}{l|}{\textbf{Value}} & \multicolumn{1}{l|}{} & \multicolumn{1}{l|}{\textbf{Parameter}} & \multicolumn{1}{l|}{\textbf{Sym.}} & \multicolumn{1}{l|}{\textbf{Value}} \\ \cline{1-3}\cline{1-3} \cline{5-7} \cline{5-7}  
			\multicolumn{1}{|l|}{Bandwidth} & \multicolumn{1}{l|}{} & \multicolumn{1}{l|}{$125$ kHz} & \multicolumn{1}{l|}{} & \multicolumn{1}{l|}{Noise PSD} & \multicolumn{1}{l|}{$N_0$} & \multicolumn{1}{l|}{$-174$ dBm/Hz} \\
			\multicolumn{1}{|l|}{Carrier Frequency} & \multicolumn{1}{l|}{} & \multicolumn{1}{l|}{$868.1$ MHz} & \multicolumn{1}{l|}{} & \multicolumn{1}{l|}{Noise Figure} & \multicolumn{1}{l|}{} & \multicolumn{1}{l|}{$6$ dBm } \\
			\multicolumn{1}{|l|}{Transmit Power} & \multicolumn{1}{l|}{} & \multicolumn{1}{l|}{$14$ dBm} & \multicolumn{1}{l|}{} & \multicolumn{1}{l|}{Duty-Cycle} & \multicolumn{1}{l|}{$\alpha$} & \multicolumn{1}{l|}{$0.33$ \%} \\
			\multicolumn{1}{|l|}{Pathloss Exponent} & \multicolumn{1}{l|}{} & \multicolumn{1}{l|}{3} & \multicolumn{1}{l|}{} & \multicolumn{1}{l|}{Coding Rate} & \multicolumn{1}{l|}{} & \multicolumn{1}{l|}{4/8} \\
			\multicolumn{1}{|l|}{Preamble} & \multicolumn{1}{l|}{} & \multicolumn{1}{l|}{8 Symbols} & \multicolumn{1}{l|}{} & \multicolumn{1}{l|}{Payload} & \multicolumn{1}{l|}{} & \multicolumn{1}{l|}{10 Bytes} \\
			\multicolumn{1}{|l|}{Clock Skew} & \multicolumn{1}{l|}{$\gamma$} & \multicolumn{1}{l|}{40 ppm} & \multicolumn{1}{l|}{} & \multicolumn{1}{l|}{Spreading Factor} & \multicolumn{1}{l|}{SF} & \multicolumn{1}{l|}{$\in \left\{7,10,12\right\}$} \\
			\multicolumn{1}{|l|}{SNR Thresholds} & \multicolumn{1}{l|}{} & \multicolumn{1}{l|}{$-\left[6,15,20\right]$ dB} & \multicolumn{1}{l|}{} & \multicolumn{1}{l|}{SIR Threshold} & \multicolumn{1}{l|}{} & \multicolumn{1}{l|}{1 dB}\\
			\multicolumn{1}{|l|}{Sync. Event Period} & \multicolumn{1}{l|}{$T_\text{SYNC}$} & \multicolumn{1}{l|}{60 s} & \multicolumn{1}{l|}{} & \multicolumn{1}{l|}{Radius} & \multicolumn{1}{l|}{$R$} & \multicolumn{1}{l|}{6 km} \\\cline{1-3} \cline{5-7} 
			&  &  &  &  &  & 
	\end{tabular}}
	\label{tb:par}
	\vspace{-15pt}
\end{table}

\subsection{Simulated Scenario}
We simulated a typical scenario in which a single gateway is located at the center of a group of LoRa end devices. The devices are randomly and uniformly distributed over a circular-shaped region of radius 6 km. We tested different configurations for device densities, spreading factors, and synchronization errors. Depending on the spreading factor, we simulated the communication scheme for a variable length of time, which varied from 20 minutes for SF 7 to 8 hours for SF 12. The  traffic in the scenario is exclusively uplink, and without any acknowledgments. Messages are generated according to a Poisson distribution with mean $\alpha \cdot \text{ToA}$. 
We first determined the baseline performance of LoRa with pure ALOHA channel access and subsequently compared it with the random access slotted communication enabled by out-of-band synchronization.   
The performance metrics used to discuss the results are normalized system throughput, data extraction rate (DER), transmission success probability, and fairness. 
We conducted simulations for both the uniform and Gaussian distributed timing errors. 
However, we present the results for the latter case only, because of the small differences observed between the two.
\subsection{Normalized System Throughput}
The first metric of interest is the normalized system throughput, measured as the fraction of channel air time used to successfully transmit messages to the gateway.
This is a system-level metric that represents how efficiently the devices utilize the channel for communication.

Fig.~\ref{fig:th} shows the normalized system throughput at different normalized traffic loads ($G$). The baseline throughput of LoRa, with its pure ALOHA protocol (LoRA~[ALOHA]), is compared to the throughput of the proposed time-slotted protocol (S-LoRa) with out-of-band synchronization.
The transmission uncertainty of the proposed time-slotted protocol, resulting from the  \mbox{FM-RDS} timing errors, requires a guard time of a few milliseconds. For a given transmission uncertainty distribution, the guard time was selected using the corresponding equation \eqref{eq:Tg_opt} or \eqref{eq:Tg_opt_n}. 
The long transmission time of messages in LoRa keeps the overhead associated with the guard time very small, especially for large SFs. 

The results in Fig.~\ref{fig:th} show that the throughput obtained from using \mbox{FM-RDS} for synchronization (S-LoRa~[FM-RDS]) is comparable to that obtainable from a very accurate time-dissemination source (S-LoRa~[Ideal]), i.e., GPS.
The low SIR threshold required by the power-capture effect of LoRa modulation results in a throughput that, compared to the pure and slotted ALOHA case without power-capture effect, is significantly higher, especially for high traffic loads. The results show that LoRa with synchronous communication can achieve high channel utilization efficiency thanks to the cumulative effect of the short vulnerability time of the synchronous communication and the low power-capture threshold of LoRa modulation.

\subsection{Data Extraction Rate and Success Probability}
The normalized system throughput offers no indication of the  reliability offered by the network. 
Subsequently, to study the communication reliability, we measured the data extraction rate (DER), defined as the ratio of received messages to transmitted messages observed in the network over a period of time~\cite{bor2016lora}.
\begin{figure}[!t]
	\centering\hspace{-2mm}
	\includegraphics[width=0.85\linewidth]{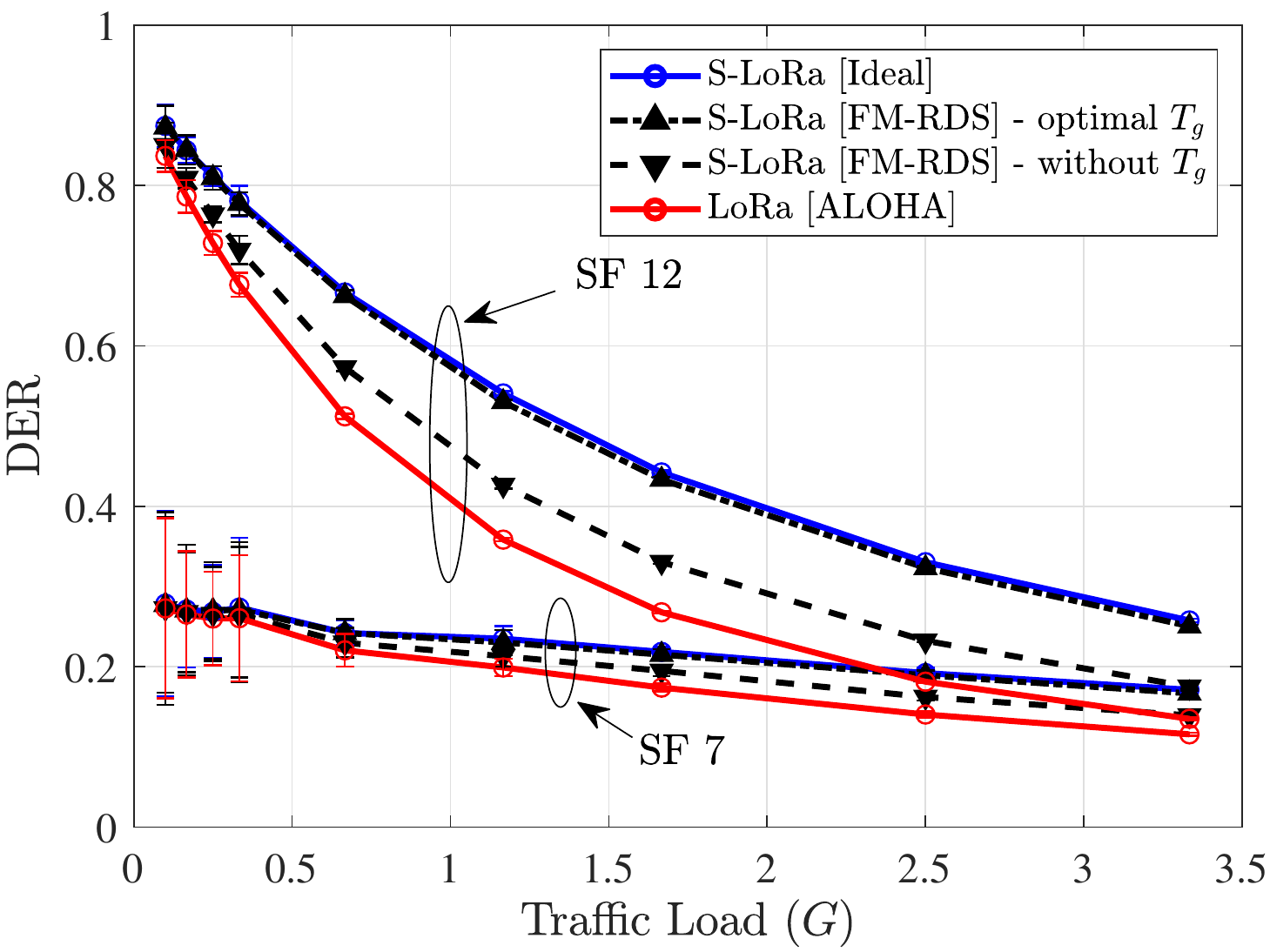}
	\caption{Data extraction rate of the network. {For the considered duty-cycle, the traffic load ($G$) is generated by a number of devices $N= \left\lbrace 30, 50, 75, 100, 200, 350, 500, 750, 1000\right\rbrace$.The error bar indicates a 50\% confidence interval.}}
	\label{fig:DER}
\end{figure}
The results, presented in Fig.~\ref{fig:DER}, show a decrease in DER for SF~7 and 12, with respect to the normalized traffic load in the network.  The lower DER of SF~7, compare to SF~12, is the result of the lower immunity to noise, which increases the probability of the gateway failing to receive messages transmitted by the devices in the given deployment region, due to the low SNR. 

\begin{figure}[!t]
	\centering
	\includegraphics[width=0.84\linewidth]{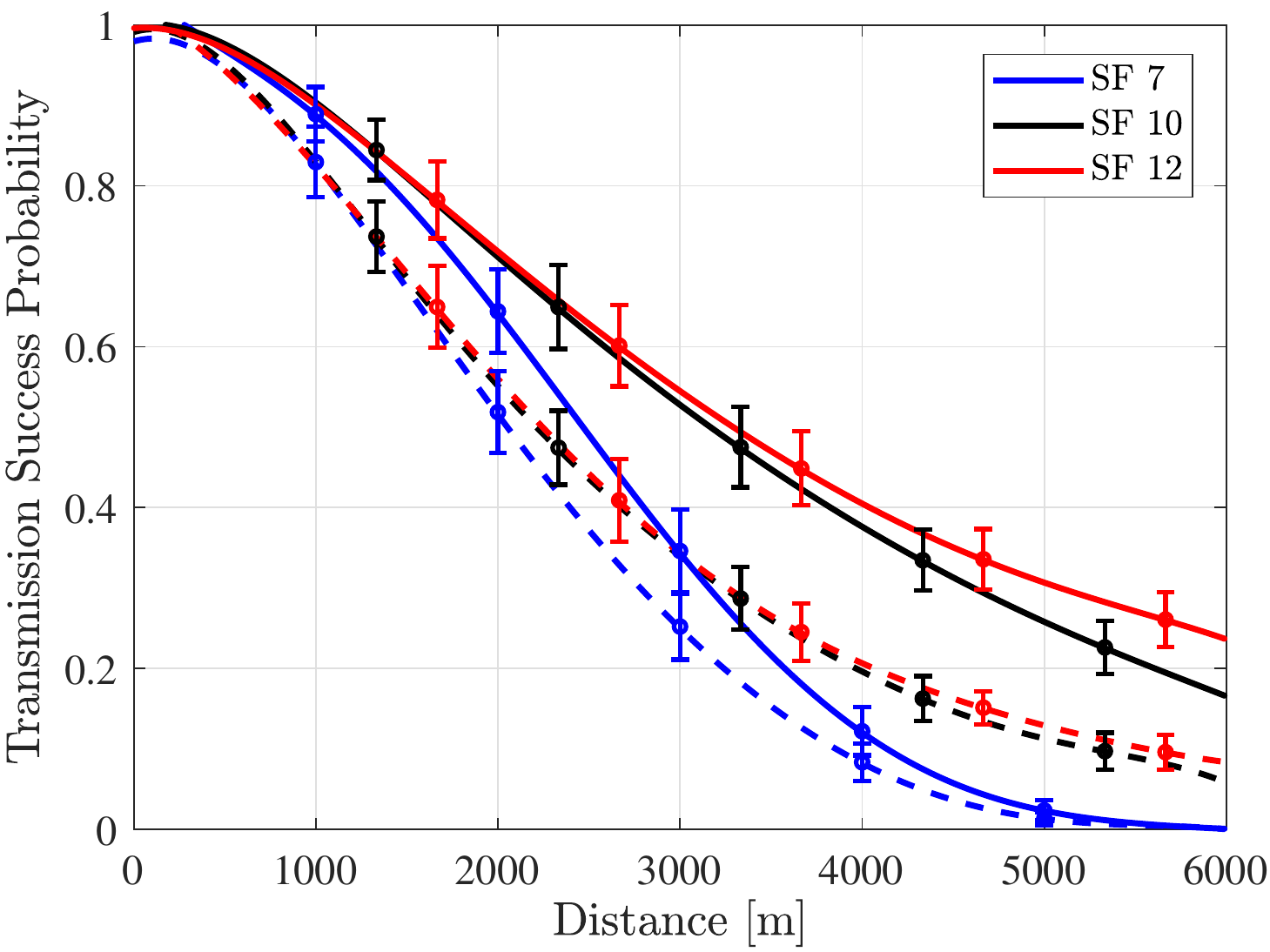}
	\caption{Success probability at different distances for LoRa~[ALOHA] (dashed lines) and S-LoRa~[FM-RDS] {with optimal $T_g$} (solid lines). The number of devices $N=500$. The error bar indicates a 50\% confidence interval.}
	\label{fig:psuc2}
 	\vspace{-10pt}
\end{figure}

We use the transmission success probability, calculated as the fraction of generated messages by each device that the gateway successfully receives, to measure the performance of individual devices in the network as a function of their distance from the gateway.  
Fig.~\ref{fig:psuc2} shows that devices which are more distant from the gateway experience a lower success probability when attempting to transmit a message. These devices are also the first to be affected by an increase in interference, given that they tend to transmit the weak messages that are more likely to be lost in the case of frame collisions. 
\subsection{Fairness}
\begin{figure*}[!ht]%
	\centering
	\subfloat[Spreading factor 7 ]{\includegraphics[width=0.332\linewidth]{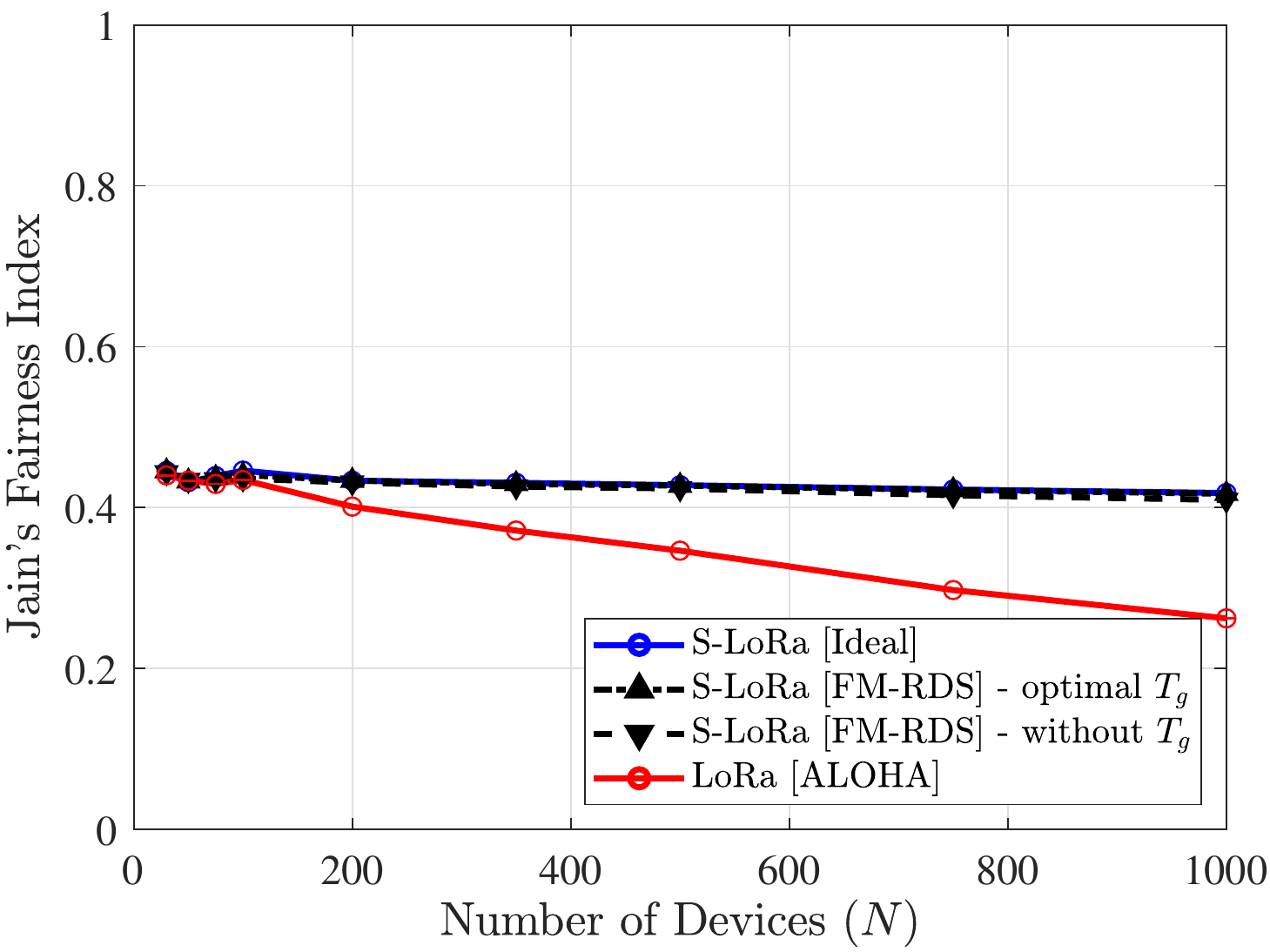}}\hfill
	\subfloat[Spreading factor 10]{\includegraphics[width=0.332\linewidth]{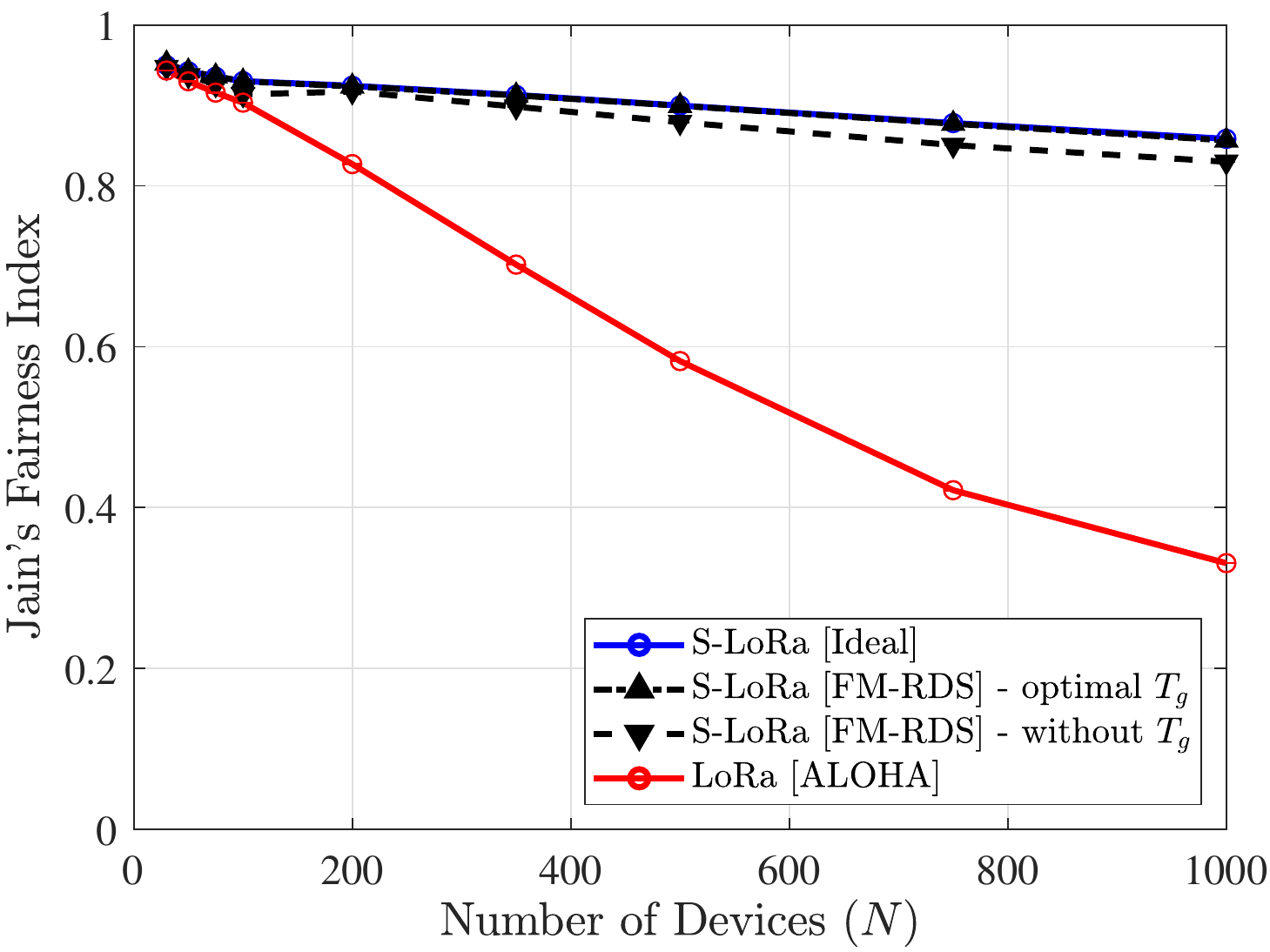}}\hfill
	\subfloat[Spreading factor 12]{\includegraphics[width=0.332\linewidth]{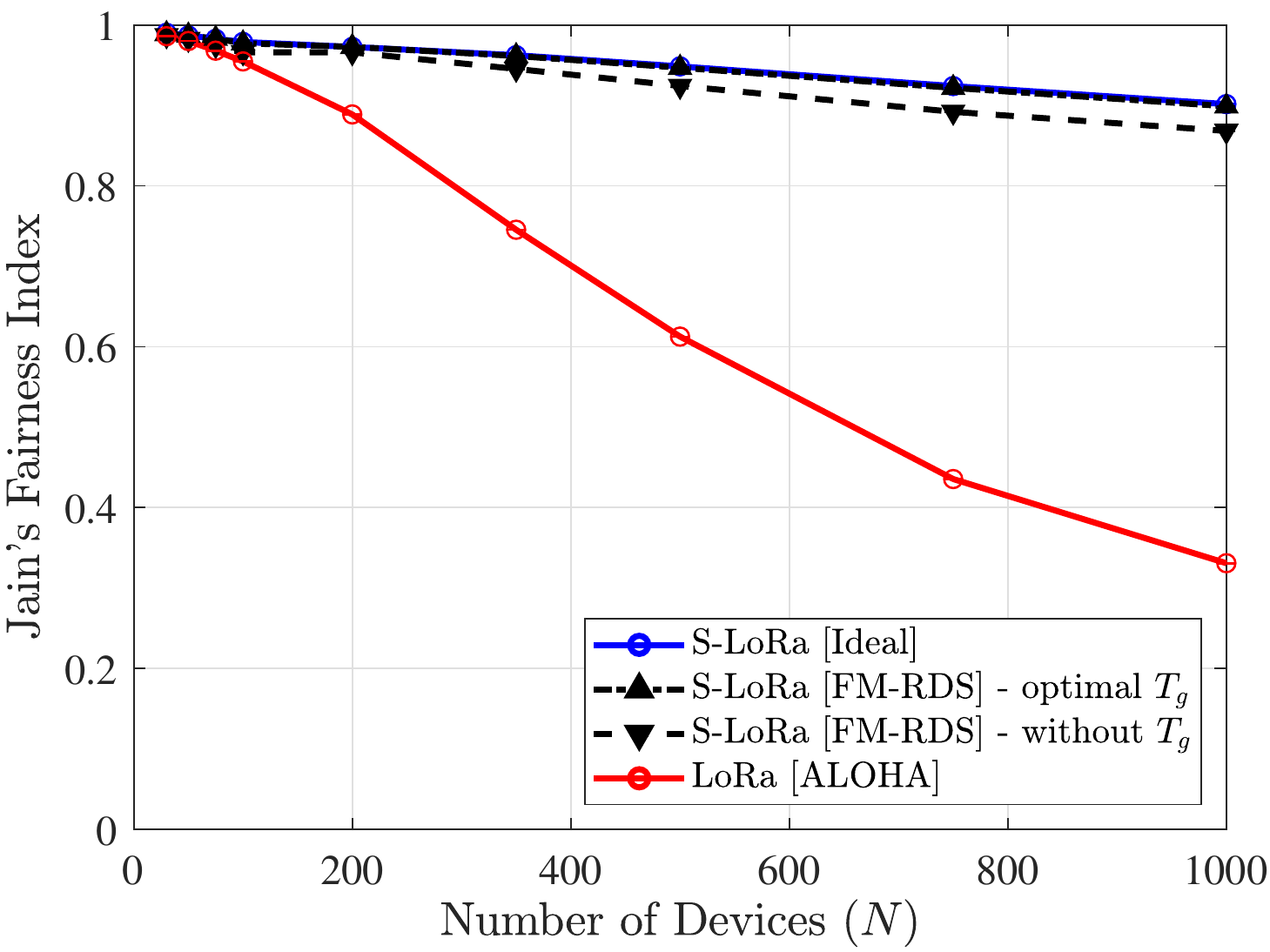}}
	\caption{Jain's fairness index of the throughput of the devices in the network.}
	\label{fig:JFI}
	\vspace{-10pt}
\end{figure*}

To measure how fairly the throughput is divided among the devices in the network, we use Jain's fairness index, which is calculated as
\begin{equation}\label{eq:JFI}
\mathcal{J}(x_1,x_2,\ldots,x_N)={\left(\sum_{k=1}^{N}x_i\right)^2} \bigg/ {\left(N\cdot \sum_{k=1}^{N}x_i^2\right)},
\end{equation}
where $x_i$ is the ratio between throughput and offered traffic load of the $i$-th device in the network.
Fig.~\ref{fig:JFI} shows Jain's fairness index for different numbers of devices $N$.
Larger SFs, due to their better coverage, tend to offer better fairness by reducing the difference in the transmission success probability between devices closer and farther from the gateway.      
The fairness index decreases as the number of devices, and hence the possible interferers, increases. The most significant decrease is measured by pure ALOHA, whereas synchronous communication is capable to handle a higher number of devices.

\section{Conclusion}
\label{Conclusion}
Our initial analysis of in-band and out-of-band synchronization revealed that, under strict duty-cycle limitations, a slotted ALOHA channel access reliant on in-band synchronization is not always capable of improving the transmission success probability of LoRaWAN's pure ALOHA. In this work, we therefore conceptualize and analyze a co-designed synchronization and random access communication scheme that uses out-of-band synchronization for resource-efficient uplink slotted communication in LoRa networks.  In the evaluation phase, we considered using \mbox{FM-RDS}, which has a time-dissemination accuracy of three orders of magnitude worse than GPS but offers a wide-area outdoor and indoor coverage, required by large-scale IoT applications. We showed that, due to the long transmission time of LoRa messages, the guard time necessary to protect against \mbox{FM-RDS} timing errors introduces a small overhead, thus only slightly reducing the throughput compared to the case in which an ideal synchronization source is used. The overall results in terms of throughput, success probability, and fairness show the advantage of using synchronous communication to overcome the current limitations of pure ALOHA in LoRaWAN. In the future, we plan to study the energy consumption and latency of the proposed mechanism.

\ifCLASSOPTIONcaptionsoff
\newpage
\fi



\bibliographystyle{IEEEtran}
\bibliography{biblio}

\end{document}